\documentclass[a4paper,11p]{article}
\usepackage{jheppub}
\usepackage{times}
\usepackage{epsfig}
\usepackage{amsmath}
\usepackage{amssymb}
\usepackage{amsfonts}
\usepackage{color}
\usepackage{slashed}
\usepackage{hyperref}
\usepackage{array}

\allowdisplaybreaks

\def\beq{\begin{equation}}
\def\eeq{\end{equation}}
\def\bea{\begin{eqnarray}}
\def\eea{\end{eqnarray}}
\def\ep{\epsilon}

\def\nn{\nonumber}
\def\Eq#1{eq.~(\ref{#1})}

\def\bq{q\hspace{-.42em}/\hspace{-.07em}}

\def\bk{k\hspace{-.42em}/\hspace{-.07em}}
\def\td#1{\tilde{\delta}(#1)}
\def\pb{\mathbf{p}}

\def\uv{{\rm UV}}

\def\r{{\rm R}}
\def\ii{\imath 0}
\def\la{\langle}
\def\ra{\rangle}
\def\vev{\la v \ra}
\def\Su#1{#1^{(+)}}

\def\Se{\widetilde{S}_{\ep}}

\def\Feynarts{{{\sc FeynArts}}}

\def\mathematica{{{\sc Mathematica}}}

\newcommand{\valencia}{IFIC, Universitat de Val\`encia-CSIC, Apt. Correus 22085, E-46071 Val\`encia, Spain}
\newcommand{\milano}{Dipartimento di Fisica, Universit\`a di Milano and INFN Sezione di Milano, I-20133 Milano, Italy}
\newcommand{\argentina}{International Center for Advanced Studies (ICAS), ECyT-UNSAM, Campus Miguelete, 25 de Mayo y Francia, (1650) Buenos Aires, Argentina.}

\begin{document}

\title{Universal four-dimensional representation of $H \to \gamma \gamma$ at two loops through the Loop-Tree Duality}

\author[a]{F\'elix Driencourt-Mangin,}
\author[a]{Germ\'an Rodrigo,}
\author[a,b,c]{Germ\'an F. R. Sborlini} 
\author[a]{and William~J.~Torres~Bobadilla}
\affiliation[a]{\valencia}
\affiliation[b]{\milano}
\affiliation[c]{\argentina}
\emailAdd{felix.dm@ific.uv.es}
\emailAdd{german.rodrigo@csic.es}
\emailAdd{german.sborlini@unimi.it}
\emailAdd{william.torres@ific.uv.es}
\preprint{IFIC/18-31 ,  TIF-UNIMI-2018-6}
\abstract{
We extend useful properties of the $H\to\gamma\gamma$ unintegrated dual amplitudes from one- to two-loop level, using the Loop-Tree Duality formalism. In particular, we show that the universality of the functional form -- regardless of the nature of the internal particle -- still holds at this order. We also present an algorithmic way to renormalise two-loop amplitudes, by locally cancelling the ultraviolet singularities at integrand level, thus allowing a full four-dimensional numerical implementation of the method. Our results are compared with analytic expressions already available in the literature, finding a perfect numerical agreement. The success of this computation plays a crucial role for the development of a fully local four-dimensional framework to compute physical observables at Next-to-Next-to Leading order and beyond. 
}

\setcounter{page}{1}         

\maketitle 
\section{Introduction}

\label{sec:Introduction}
The calculation of observables the physics the LHC delivers has
nowadays been improved with several techniques to make predictions
at the Next-to-Leading order (NLO) accuracy. In general, it is aimed
at reducing the scale uncertainties from ${\cal O}(10 \, \%)$ to ${\cal O}(1 \, \%)$, and to even less for some specific processes such as Drell-Yan. In
order to provide these observables, we
rely our prediction on the calculations of scattering amplitudes through perturbation theory.
For the calculation of the latter at NLO, apart from evaluating
Feynman integrals, we also need to deal with the evaluation of integrals in
the momentum space. At one-loop level, the basis of integrals is known
and their evaluation has been implemented in several codes (for instance, in refs.~\cite{vanHameren:2010cp,Carrazza:2016gav}).
Nevertheless, the evaluation of multi-loop integrals still remains a work in progress~\cite{Borowka:2015mxa,Smirnov:2015mct}. 
On top of it, depending on the process under
consideration, there may appear infrared (IR) and ultraviolet (UV) 
singularities that are canceled out by adding real corrections and
proper counter-terms. 

The two-loop QCD corrections to the decay process $H \to \gamma \gamma$ have been first evaluated
in the heavy-top limit~\cite{Harlander:2005rq,Zheng:1990qa,Djouadi:1990aj,Dawson:1992cy}
and with the full top-mass dependence~\cite{Fleischer:2004vb,Aglietti:2006tp}.
The two-loop electroweak corrections have been investigated in refs.~\cite{Actis:2008ts,Passarino:2007fp,Degrassi:2005mc,Fugel:2004ug}.
Combining the two-loop QCD and electroweak corrections, it is possible to observe a nearly complete cancellation between these two contributions for
$M_{H}=126$ GeV~\cite{Maierhofer:2012vv}. At Next-to-Next-to-Leading
order (NNLO) the non-singlet~\cite{Steinhauser:1996wy} and singlet
QCD contributions~\cite{Maierhofer:2012vv} have been calculated in
the heavy top quark limit.

In view of the enormous success of the Standard Model (SM) of particle physics with the detection
of the Higgs boson, new directions have been taken to discuss in more
details the consequences of this discovery. In particular, from the phenomenological point
of view, the background of the experiment has to be removed. Hence,
QCD predictions up to the Next-to-Next-to-Next-to-Leading order (N$^{3}$LO) have been provided in an effective theory~\cite{Anastasiou:2016cez}.
Also, it has been shown that the mixed effects of QCD-electroweak
contribution to the amplitude are relevant~\cite{Bonetti:2018ukf}. 

Nevertheless, hidden mathematical properties of the amplitudes $gg\to H$
and $H\to\gamma\gamma$ have been extensively studied in the full
theory at Leading order (LO) in ref.~\cite{Driencourt-Mangin:2017gop},
in which, it was showed that these amplitudes exhibited remarkable
properties when computed using the Loop-Tree Duality (LTD) theorem~\cite{Catani:2008xa,Bierenbaum:2010cy,Bierenbaum:2012th}.
The dual contributions we have obtained for different internal particles
-- charged electroweak gauge bosons, massive fermions and charged scalars
-- featured the very same functional forms, and could be written
in a universal way using scalar parameters depending only on the space-time
dimension ($d$), and the mass of the particles involved in the process.
We also obtained a pure four-dimensional ($d=4$) representation of the renormalised amplitude and recovered the well-known results found in the
literature~\cite{Wilczek:1977zn,Georgi:1977gs,Rizzo:1979mf,Ellis:1975ap,Ioffe:1976sd,Shifman:1979eb}.

In this manuscript we push the computation further by considering
the $H\to\gamma\gamma$ process at two-loop level, and show that the
above-mentioned properties are still present. To this end, we make
use of the LTD theorem, which converts loop integrals into phase-space ones. In order
to provide the renormalised amplitude, we perform a local UV renormalisation
that leads to a finite integrand in four space-time dimensions. This algorithm is based on the refinement of the \emph{expansion around the UV propagator} \cite{Becker:2010ng,Sborlini:2016gbr,Sborlini:2016hat} to account for the different singular behaviours of the internal loop momenta in the UV region.
Furthermore, since this amplitude is IR safe, we can directly treat the virtual integrand
in four dimensions. We remark that the calculation of this amplitude
is the first two-loop application to a physical process done through
LTD, and it is computed below the mass threshold limit in the $\overline{\text{MS}}$
renormalisation scheme. We note that for individual diagrams, unphysical threshold
singularities appear but they cancel among themselves when the full amplitude is considered. 

In the same spirit of the universality that these amplitudes exhibit at LO, we consider as internal particles charged scalars and top quarks. 
While we only consider QED corrections, they can be straightforwardly promoted
to QCD ones by replacing the couplings accordingly. We compare our results with known results~\cite{Fleischer:2004vb,Aglietti:2006tp}
finding full agreement. 

With this paper we verify that the LTD approach holds also at multi-loop
level and, therefore, N$^{k}$LO predictions involving virtual amplitudes
can be achieved by means of the former. Additionally, we remark that the traditional
approach based on the use of integration-by-parts identities~\cite{Chetyrkin:1981qh,Laporta:2001dd} is not
needed to evaluate the actual amplitude. In fact, we overcome the calculation of the latter making our procedure much lighter as we shall describe here. 

The paper is organised as follows. In section \ref{sec:twoloops}, we recall the basics of the LTD formalism, at one- and two-loop level. We introduce there the notations used in this paper, and provide the master formula for obtaining the dual representation of a two-loop Feynman integral. In section \ref{sec:reduction}, we sketch the algorithm to algebraically reduce the integrand-level expressions of two-loop dual amplitudes, and rewrite every scalar product involved in terms of denominators. We provide the tensor structure of the $H\to\gamma\gamma$ amplitude in section \ref{sec:tensorproj}, and briefly recall the one-loop result obtained in ref. \cite{Driencourt-Mangin:2017gop}. Then, in section \ref{sec:hgg}, we collect and write the universal coefficients involved in the universal structure of the two-loop dual expressions. We discuss in section \ref{sec:cancellation} the cancellations of unphysical threshold singularities that appear among the dual contributions, and we explicitly show how they occur. In section \ref{sec:UVren}, we discuss an algorithmic approach to locally renormalise two-loop amplitudes within the LTD formalism. In particular, we focus on the determination of the scheme-fixing parameters in the $\overline{\text{MS}}$ scheme. In section \ref{sec:numint}, we present our numerical results and show there is a complete agreement with the analytical expressions. We draw our conclusions and discuss future directions of this work in section \ref{sec:conclusions}.

\bigskip Algebraic manipulations have been carried out by using an
in-house implementation of LTD which is based on the \textsc{Mathematica}
packages \textsc{FeynArts}~\cite{Hahn:2000kx} and \textsc{FeynCalc}~\cite{Mertig:1990an, Shtabovenko:2016sxi}.

\section{The Loop-Tree Duality at two loops}
\label{sec:twoloops}

The Loop-Tree Duality (LTD) theorem~\cite{Catani:2008xa,Bierenbaum:2010cy,Bierenbaum:2012th} transforms any loop integral or loop scattering 
amplitude into a sum of tree-level  like objects that are constructed by setting on shell a number of internal loop propagators
equal to the number of loops. Explicitly, LTD is realised by modifying the ${\imath 0}$ prescription of the Feynman propagators
that remain off shell 
\begin{equation}
G_F(q_j) = \frac{1}{q_j^2-m_j^2+\imath 0} \qquad \to \qquad G_D(q_i;q_j) = \left. \frac{1}{q_j^2-m_j^2-\imath 0 \, \eta k_{ji}} \right|_{G_F(q_i) ~ {\rm on-shell}}~,
\end{equation}
with $k_{ji} = q_j-q_i$, and $\eta^\mu$ an arbitrary future-like vector. The most convenient choice is $\eta^\mu=(1,\bf{0})$, which is equivalent
to integrate out the loop energy components of the loop momenta through the Cauchy residue theorem. The left-over integration is
then restricted to the Euclidean space of the  loop three-momenta. The dual prescription can hence be either $-\imath 0 \, \eta k_{ji}=-\imath 0$ 
for some dual propagators or $-\imath 0 \, \eta k_{ji}=+\imath 0$ for the others, since indeed only the sign matters. In fact, this prescription encodes in a compact and elegant way the contribution of the multiple cuts that are 
introduced by the Feynman tree theorem~\cite{Feynman:1963ax}. The on-shell condition is given by 
$\tilde \delta(q_i) = \imath \, 2 \pi \, \theta(q_{i,0}) \, \delta(q_i^2-m_i^2)$, and determines that the
loop integration is restricted to the positive energy modes, $q_{i,0}>0$, of the on-shell hyperboloids (light-cones for massless particles)
of the internal propagators. We also introduce the short-hand notation for the loop integration measure in $d$ dimensions,
\beq
\int_{\ell_i} \bullet =-\imath\, \mu^{4-d}\int \frac{d^d \ell_i}{(2\pi)^{d}} \; \bullet~,
\eeq
where $\mu$ is an arbitrary mass scale to compensate the extra dimensions generated by $d$-dimensional integration measure. In the following, we use $d=d_s=4-2\epsilon$ according to the convention 
of ref.~\cite{Gnendiger:2017pys}.

In order to generalise LTD to higher orders, we introduce the following functions~\cite{Bierenbaum:2010cy}
\beq
G_F(\alpha_k) = \prod_{i\in \alpha_k} G_F(q_i)~, \qquad
G_D(\alpha_k) = \sum_{i\in \alpha_k} \td{q_i} \,
\prod_{\substack{j\in \alpha_k \\ j\ne i}} G_D(q_i;q_j)~, 
\label{eq:fdual}
\eeq
where $\alpha_k$ labels all the propagators, Feynman or dual, of a given subset. 
An interesting identity fulfilled by these functions is the following 
\beq
\label{eq:split}
G_D(\alpha_i \cup \alpha_j) = G_D(\alpha_i) \, G_D(\alpha_j) + G_D(\alpha_i) \, G_F(\alpha_j) + G_F(\alpha_i) \, G_D(\alpha_j)~,
\eeq
involving the union of two subsets $\alpha_i$ and $\alpha_j$. 
These are all the ingredients necessary to iteratively extend LTD to multi-loop level. 
For example, at one loop, the Feynman and the dual representations of a $N$-leg scattering amplitude are 
\beq
{\cal A}^{(1)}_N = \int_{\ell_1} {\cal N}(\ell_1,\{p_i\}_N) \, G_F(\alpha_1) = - \int_{\ell_1} {\cal N}(\ell_1,\{p_i\}_N) \, \otimes \, G_D(\alpha_1)~,
\label{eq:dualoneloop}
\eeq
respectively, where  ${\cal N}(\ell_1,\{p_i\}_N)$ is the numerator that depends on the 
loop momentum $\ell_1$ and the four-momenta of the $N$ external partons $\{p_i\}_N$. 
In the absence of multiple powers of the Feynman propagators, the numerator is not altered by 
the application of the Cauchy theorem. However, the calculation of the residues of multiple poles to obtain the corresponding LTD representation 
requires the participation of the numerator. This is represented in \Eq{eq:dualoneloop} by the symbol $\otimes$. 

\begin{figure}[th]
\begin{center}
\includegraphics{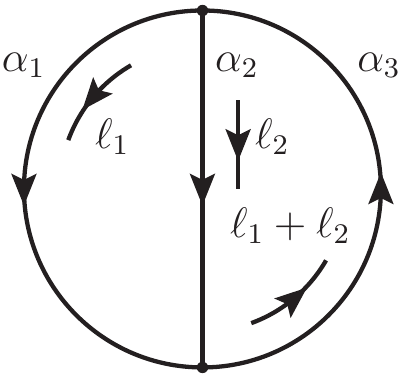}
\caption{Momentum flow of a two-loop Feynman diagram. An arbitrary number of external legs (not shown) are attached 
to each loop line $\alpha_i$.
\label{fig:alphas}}
\end{center}
\end{figure}

At two-loop level, all the internal propagators can be classified into three different subsets (e.g. those depending on $\ell_1$, $\ell_2$ and their sum $\ell_{12}=\ell_1+\ell_2$,
as shown in figure~\ref{fig:alphas}). Starting from the Feynman representation of a two-loop scattering amplitude
\beq
\label{Ln}
{\cal A}^{(2)}_N =
\int_{\ell_1} \, \int_{\ell_2} {\cal N} (\ell_1, \ell_2,\{p_i\}_N) \, G_F(\alpha_1\cup \alpha_2 \cup \alpha_3)~, 
\eeq
by applying LTD to one of the loops (\Eq{eq:dualoneloop}), we obtain in a first step:
\beq
{\cal A}^{(2)}_N =
- \int_{\ell_1} \, \int_{\ell_2} {\cal N} (\ell_1, \ell_2,\{p_i\}_N) \, G_F(\alpha_1) \, G_D(\alpha_2 \cup \alpha_3)~.
\label{eq:nodouble}
\eeq
Before applying LTD to the second loop, it is necessary to use \Eq{eq:split} to express the dual function $G_D(\alpha_2 \cup \alpha_3)$
in a suitable form. The identity in \Eq{eq:split} splits the dual integrand into a first term that contains two dual functions -- and therefore two internal lines on shell -- and two more terms with a single dual function and Feynman propagators involving the other two sets of propagators, to which we can recursively apply LTD. The final dual representation of the two-loop amplitude in \Eq{Ln} is
\bea
{\cal A}^{(2)}_N  &=& 
\int_{\ell_1} \, \int_{\ell_2} {\cal N} (\ell_1, \ell_2,\{p_i\}_N) \, \otimes \, \bigg\{  G_D(\alpha_2) \, G_D(\alpha_1 \cup \alpha_3) \nn \\ &+& 
G_D(-\alpha_2 \cup \alpha_1) \, G_D(\alpha_3)  
-  G_F(\alpha_1) \, G_D(\alpha_2) \, G_D(\alpha_3) \bigg\}~.
\label{twoloopduality}
\eea 
In \Eq{twoloopduality}, it is necessary to take into account that the momentum flow in the loop formed by the union 
of $\alpha_1$ and $\alpha_2$ occurs in opposite directions. Therefore, it is compulsory to change the direction 
of the momentum flow in one of the two sets. This is represented by adding a sign in front of e.g. $\alpha_2$, explicitly we have
\beq
\int_{\ell_1} \int_{\ell_2} \, G_F(\alpha_1) \, G_F(\alpha_2)  = - \int_{\ell_1} \int_{\ell_2} \, G_D(-\alpha_2 \cup \alpha_1)~.
\eeq
Changing the momentum flow is equivalent to select the negative energy modes. For the internal momenta 
in the set $\alpha_2$, this means 
\beq
\td{-q_j} = \frac{\imath \, \pi}{q_{j,0}^{(+)}} \, \delta(q_{j,0}+q_{j,0}^{(+)})~, \qquad j \in \alpha_2~.
\eeq

The dual representation gets its simplest form if the Feynman representation contains only single powers of the Feynman propagators. 
This restriction cannot be avoided anymore at two-loops where, for example, self-energy insertions in internal lines 
lead automatically to double powers of one propagator. However, all the double poles can be included with a clever labelling 
of the internal momenta in the set $\alpha_1$, exclusively, which is not integrated in the first instance. 
Therefore, we have assumed that the numerator in \Eq{eq:nodouble} is not affected by the application of LTD. 
The final dual representation in \Eq{twoloopduality}
depends, in general, on the explicit form of the numerator. Again, this is represented by the symbol $\otimes$.

The number of independent double cuts in \Eq{twoloopduality} per Feynman diagram is 
\beq
{\rm N}(\alpha_2\times (\alpha_1+\alpha_3) + (\alpha_1+\alpha_2)\times \alpha_3)~,
\eeq
where ${\rm N}$ counts the number of propagators in each set. 
Therefore, it is convenient to have $\alpha_2$ as the set with the smallest number of propagators. 
For planar diagrams, the set $\alpha_2$ contains one single propagator. 

It is interesting to note that although the integration over the loop three-momenta is unrestricted, after analysing the singular behaviour of the loop integrand one realises that thanks to a partial cancellation of singularities among different dual components, all the physical threshold and 
IR singularities remain confined to a compact region of the loop three-momentum~\cite{Buchta:2014dfa,Buchta:2015wna}.
This relevant fact allows to construct mappings between the virtual and real kinematics,
which are based on the factorisation properties of QCD, to implement the summation over degenerate soft and collinear states
for physical observables in the Four-Dimensional Unsubtraction (FDU) formalism~\cite{Hernandez-Pinto:2015ysa,Sborlini:2016gbr,Sborlini:2016hat}. This framework, however, is not going to be needed in the following. This is because, as we shall see in section~\ref{sec:cancellation} where we analyse the cancellation of singularities at two loops, no infrared singularities remain when considering the full amplitude.

\section{Algebraic reduction of two-loop dual amplitudes}
\label{sec:reduction}

In order to make the two-loop expressions more compact, we  perform an algebraic reduction of the dual amplitudes to dual integrals that involve 
both positive and negative powers of dual propagator denominators. We analyse only the case of planar diagrams that are those that appear in 
the practical example that we present. 


Let us first consider scattering amplitude with $N$ external legs and ordered external momenta\linebreak $\{p_1,p_2,\ldots,p_N\}$. At one-loop, we have $N$ different propagators and $N-1$ independent scalar products $\ell_1\cdot p_i$ (indeed, because of momentum conservation, $\ell_1\cdot p_N=-\sum_{i=1}^{N-1} \ell_1 \cdot p_i$, with $\ell_1$ the loop four-momentum). In the Feynman representation, the propagators are quadratic in $\ell_1$, while in LTD the dual propagators are linear. In both formalisms, however,
it is possible to write numerators in terms of propagators diagram by diagram. 

\begin{figure}[t]
	\begin{center}
		\includegraphics[width=0.65\textwidth]{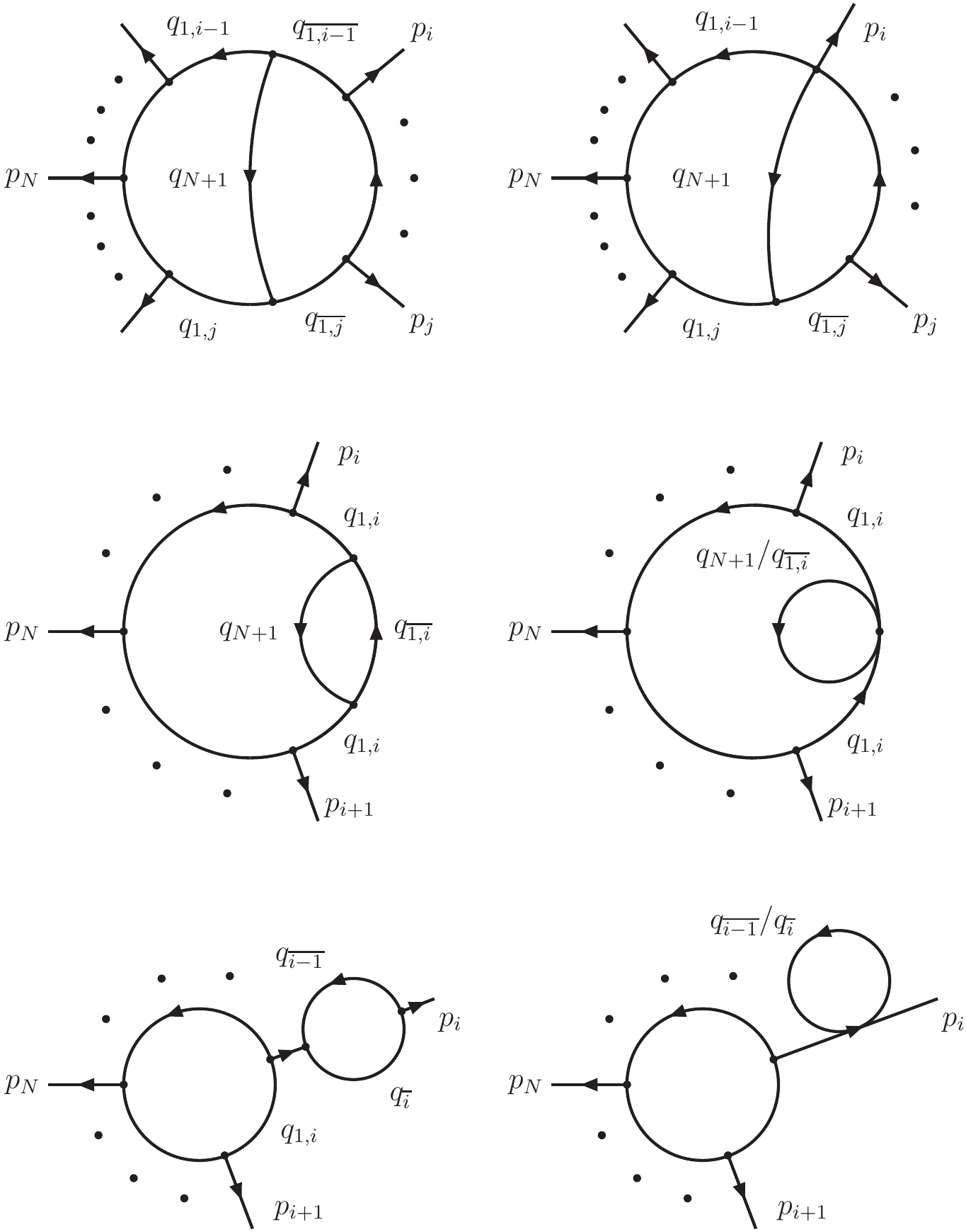}
		\caption{Assignment of momenta in two-loop planar diagrams. 
			\label{fig:planar}}
	\end{center}
\end{figure}

Now, we consider the set of two-loop planar Feynman diagrams constructed from the ordered one-loop seed diagram -- that is, all the two-loop diagrams that have one loop line involving one single propagator (see figure~\ref{fig:planar} for the assignment of momenta). These planar two-loop Feynman diagrams can be constructed from the sets of propagators 
\begin{equation}
\alpha_1 = \{ q_1, q_{12}, \ldots, q_{1,N}\}~,  \qquad
\alpha_2 = \{q_{N+1}\}~, \qquad
\alpha_3 = \{q_{\overline{1}}, q_{\overline{12}}, \ldots, q_{\overline{1,N}}\}~,
\end{equation}
with $q_{i,j} = \ell_1 + p_{i,j}$, $q_{N+1}=\ell_2$, $q_{\overline{i,j}} = \ell_{12} + p_{i,j}$ and $p_{i,j}=p_i+p_{i+1}+\dots+p_j$. These propagators are constructed by attaching $\ell_2$ in all possible ways while keeping the same ordering of the external momenta in the loop formed by the other two loop lines $\alpha_1$ and $\alpha_3$. 
This sums to $\text{N}(\alpha_1+\alpha_2+\alpha_3)=2N+1$ possible propagators.
If there are only three-point interactions, each Feynman diagram contains $N+3$ propagators from these sets. 
If there are $V_4$ four-point interaction vertices, each individual Feynman diagram contains only $N+3-V_4$ propagators.

Because of momentum conservation, there are $2N-1$ independent scalar products that are
\beq
 \{\ell_1\cdot p_i, \ell_2\cdot p_j, \ell_1\cdot\ell_2~~|~~ i,j\in \{1,\ldots, N-1\} \}~.
\eeq
In LTD at two loops, two internal particles are set on shell, which means that only $2N-1$ dual propagators remain  for a given double cut. Moreover, dual propagators are linear in each of the loop momenta, and the dual numerators do not involve squared loop-momenta. Therefore, for each double cut, and considering all the Feynman diagrams with the same ordering of the external particles, it is possible to rewrite all the scalar products involved (and thus the numerators) in terms of dual propagators, and this in a unique way. 
Due to the assignment of the loop momenta, all the required irreducible scalar products (ISP) are automatically introduced. This is because the set of Feynman diagrams contains all the necessary propagators to perform the complete algebraic reduction.

The algebraic reduction of a planar two-loop dual amplitude with $N$ external legs,
and at most squared propagators in one single loop line, leads to 
\bea
{\cal A}^{(2)}_N &=& \int_{\ell_1} \int_{\ell_2} {\cal N}(\ell_1,\ell_2,\{p_i\}_N) \, G_F(\alpha_1 \cup \alpha_2 \cup \alpha_3) + {\rm perm.} \nn \\
&=&  \int_{\ell_1} \int_{\ell_2} \sum_{j,k} 
\left[ \frac{c_{a_0; a_1, \ldots, a_{2N-1}} (\{p_i\}_N) }{(q_{j,0}^{(+)})^{a_0} (d_{i_1})^{a_1} (d_{i_2})^{a_2} \cdots (d_{i_{2N-1}})^{a_{2N-1}}} \right] \, \td{q_j, q_k}
+ {\rm perm.} 
\eea
with\footnote{At this point is is not necessary to make the distinction, because...} $d_{i_l}  = (G_D(q_j;q_{i_l}))^{-1}$ or $d_{i_l}  = (G_D(q_k;q_{i_l}))^{-1}$, and 
\beq
\sum_{i=1}^{2N-1} a_i \le N+1~, \qquad a_{0} = \{2,1,0\}~.
\eeq
The scalar coefficients $c_{a_0; a_1, \ldots, a_{2N-1}}$ depend only on the external momenta, and are not necessarily independent. 
Our purpose is to rearrange the expressions for the dual amplitudes  in order to obtain the minimal set of independent coefficients
$c_{a_0; a_1, \ldots, a_{2N-1}}$ .  Another relevant issue to obtain  the most compact integrand expressions is
to label the internal momenta in the most symmetric way. In figure~\ref{fig:planar}, we show the assignments that we use
in the most general case for planar two-loop diagrams. Computer algebra programs for the automatic generation of 
two-loop amplitudes, like \Feynarts~\cite{Hahn:2000kx}, might use a different criteria
which require a relabelling of the internal propagators to achieve the most suitable assignment. 
In the next sections, we shall illustrate the full procedure with the benchmark amplitude $H\to\gamma\gamma$.

\section{Tensor projection and $H\to \gamma\gamma$ at one loop}
\label{sec:tensorproj}

The scattering amplitude describing the Higgs boson decay to two photons is given by 
\beq
\label{VirtualAmplitude}
|{\cal M}_{H \to \gamma\gamma} \rangle= \imath \, e^2\, \left( \sum_{f=\{\phi,t,W\}} e^2_f \, N_C^f \, {\cal A}_{\mu\nu}^{(f)} \right) 
(\varepsilon^\mu(p_1))^*(\varepsilon^\nu(p_2))^*\,,
\eeq
with $e$ the electromagnetic coupling, $e_f$ (respectively $N_C^f$) the electric charge (respectively the number of colours) of the virtual
particle $f$, and $\varepsilon(p_i)$ the polarisations vectors of the external photons. Here, we assume that the two photons are coupled to the same
flavour. We are interested in the QED corrections at two-loop level with $f=\{\phi,t,W\}$. The tensor amplitude ${\cal A}^{(f)}_{\mu\nu}$
fulfils the perturbative expansion 
\beq
{\cal A}^{(f)}_{\mu\nu}  = {\cal A}^{(1,f)}_{\mu\nu} +  (e\, e_f )^2 \, {\cal A}^{(2,f)}_{\mu\nu} + {\cal O}(e^4)~,
\eeq
where ${\cal A}^{(1,f)}_{\mu\nu}$ is the one-loop amplitude, and ${\cal A}^{(2,f)}_{\mu\nu}$ is the two-loop QED correction.
The tensor amplitudes $\mathcal{A}_{\mu\nu}^{(L, f)}$ can be decomposed through Lorentz and gauge invariance as
\beq
{\cal A}_{\mu\nu}^{(L,f)}=\sum\limits_{i=1}^6 {\cal A}_i^{(L ,f)}\, T_{i,\mu\nu}\,,
\eeq
in terms of the tensor basis
\begin{equation}
T_i^{\mu\nu}=\left\{g^{\mu\nu}-\frac{2p_2^{\mu}\,p_1^{\nu}}{s_{12}},g^{\mu\nu},\frac{2p_1^{\mu}\,p_2^{\nu}}{s_{12}},\frac{2p_1^{\mu}\,p_1^{\nu}}{s_{12}},\frac{2p_2^{\mu}\,p_2^{\nu}}{s_{12}}, \epsilon^{\mu\nu\sigma\rho} \frac{2p_{1,\sigma}\,p_{2,\rho}}{s_{12}} \right\}\,.
\end{equation}
The tensor structure $T_6^{\mu\nu}$ may appear for the first time at two-loop, because of the potential presence of a $\gamma^5$, but its interference with the one-loop amplitude vanishes. As in ref.~\cite{Driencourt-Mangin:2017gop}, we use the projectors
\beq
\label{Projectors}
P_1^{\mu\nu}=\frac{1}{d-2}\left(g^{\mu\nu}-\frac{2p_2^\mu\,p_1^\nu}{s_{12}}-(d-1)\frac{2p_1^\mu\,p_2^\nu}{s_{12}}\right)
\qquad \text{and} \qquad P_2^{\mu\nu}=\frac{2p_1^\mu\,p_2^\nu}{s_{12}}\,
\eeq
to extract the scalar amplitudes $\mathcal{A}_1^{(L,f)}$ and $\mathcal{A}_2^{(L,f)}$. There is no need to compute $\mathcal{A}_i^{(L,f)}$, for $i\in\{3,4,5\}$, as they vanish after contracting with the polarisation vectors, and therefore do not contribute to the scattering amplitude for on-shell photons. Moreover, because of gauge invariance, $\mathcal{A}_2^{(L,f)}$ is expected to vanish after integration, which leaves $\mathcal{A}_1^{(L,f)}$ as the only relevant physical term. It is still interesting to consider and get integrand expressions for $\mathcal{A}_2^{(L,f)}$, though, as it can be used to simplify expressions.

The dual representations of the one-loop amplitudes $\mathcal{A}_1^{(1,f)}$ and $\mathcal{A}_2^{(1,f)}$ were calculated in 
ref.~\cite{Driencourt-Mangin:2017gop}  in terms of the global factor $g_f$ and scalar coefficients $c_i^{(f)}$, which have the form 
$c_i^{(f)} = c_{i,0}^{(f)} + r_f\, c_{i,1}^{(f)}$ with $r_f = s_{12}/M_f^2$ and $f=\{\phi, t, W\}$. For the three different virtual particles that we considered, 
these coefficients are given by
\bea
&& g_f = \frac{2\, M_f^2}{\vev \, s_{12}}~,  \quad  c_{1,0}^{(f)} = \left( \frac{4}{d-2},-\frac{8}{d-2},\frac{4(d-1)}{d-2} \right) ~, \quad c_{1,1}^{(f)} = \left(0,1,\frac{2(5-2d)}{d-2} \right)~, \nn \\ 
&& c_{3,0}^{(f)} = (d-2) \, c_{1,0}^{(f)}~,  \quad c_{3,1}^{(f)} = 0~, \quad c_{23,0}^{(f)} = (d-4)\frac{c_{1,0}^{(f)}}{2}~, \quad c_{23,1}^{(f)} = \left(0,0, \frac{d-4}{d-2} \right)~.
\label{eq:coeffoneloop}
\eea
The vanishing one-loop amplitude ${\cal A}_2^{(1,f)}$ is given by 
\beq
{\cal A}_2^{(1,f)} = g_f\, \frac{c_3^{(f)}}{2} \, \int_{\ell_1} \, \left( \td{q_1}  + \td{q_2} - 2 \, \td{\ell_1} \right) = 0~,
\label{universal2}
\eeq
where $q_i = \ell_1 + p_i$, with $i=1,2$. The on-shell loop energies are given by
\bea
&& q_{i,0}^{(+)} = \sqrt{(\boldsymbol{\ell}_1+\pb_i)^2+M_f^2}~, \quad
\ell_{1,0}^{(+)}  = \sqrt{\boldsymbol{\ell}_1^2+M_f^2}~.
\label{eq:eonshell}
\eea
As stated above, we know that $\mathcal{A}_2^{(L,f)}=0$ after integration due to gauge invariance. Therefore, we can use this feature to simplify the expression for ${\cal A}_1^{(L,f)}= P_1^{\mu\nu} \, {\cal A}_{\mu \nu}^{(L,f)}$. Already at one loop ($L=1$), we showed that the following transformation~\cite{Driencourt-Mangin:2017gop} 
\begin{equation}\label{A1toA}
\mathcal{A}^{(L,f)} \to \mathcal{A}_1^{(L,f)}-\frac{2 \, c_2^{(f)}}{c_3^{(f)}}\mathcal{A}_2^{(L,f)}~,
\end{equation}
with $c_2^{(f)} = c_{23}^{(f)} - c_3^{(f)}$, and 
\beq
\frac{2 \, c_2^{(\phi)}}{c_3^{(\phi)}} = \frac{2 \, c_2^{(t)}}{c_3^{(t)}} = - \frac{d}{d-2}~,
\eeq
reduces the number of necessary independent scalar coefficients $c_i^{(f)}$ to describe ${\cal A}_1^{(1,f)}$ from three to two. Notice that while they are labelled differently because they do not have the same integrand-level expressions, we have $\mathcal{A}^{(L,f)}=\mathcal{A}_1^{(L,f)}$ after integration\footnote{Note that ${\cal A}_2^{(2,f)}$ integrates to 0 in $d$ dimensions, whereas in four space-time dimensions it is only the case after local renormalisation.}. For the complete expression of the amplitude, we obtain
\bea
{\cal A}^{(1,f)} &=& \, g_f\, s_{12}\, \int_{\ell_1} \td{\ell_1} \bigg[
\bigg( \frac{\ell_{1,0}^{(+)}}{q_{1,0}^{(+)}} +  \frac{(2\ell_1\cdot p_{12})^2}{s_{12}^2-(2\ell_1\cdot p_{12}-\ii)^2} \bigg) 
\frac{M_f^2}{(2\, \ell_1 \cdot p_1)(2\, \ell_1 \cdot p_2)}\, c_1^{(f)}  \nn \\
&+& \frac{s_{12}}{s_{12}^2-(2\ell_1\cdot p_{12}-\ii)^2}\, c_{23}^{(f)} \bigg] + \{p_1\leftrightarrow p_2\}~.
\label{eq:compact}
\eea
It is remarkable to note that the dependence on the nature of the internal particle appears in \Eq{universal2} and \Eq{eq:compact}
only through the scalar coefficients $c_i^{(f)}$, defined in \Eq{eq:coeffoneloop}. 

Although ${\cal A}^{(1,f)}$ is UV finite, because there is no direct coupling of the Higgs boson to photons at tree level, 
its integrand expression still exhibits a local UV behaviour. For that reason, we defined in ref.~\cite{Driencourt-Mangin:2017gop}
the UV counter-term  
\beq
{\cal A}_{\uv}^{(1,f)} = -  g_f  \, s_{12} \, \int_{\ell_1}  \frac{\td{q_{\uv}}}{2 (q_{\uv,0}^{(+)})^2}
\Bigg( 1 + \frac{1}{(q_{\uv,0}^{(+)})^2} \frac{3\, \mu_\uv^2}{d-4} \Bigg)\, c_{23}^{(f)} = 0~,
\label{eq:A1dualgauge}
\eeq 
with $q_{\uv,0}^{(+)} = \sqrt{\boldsymbol{\ell}_1^2+\mu_\uv^2}$. 
The UV counter-term in \Eq{eq:A1dualgauge} integrates to zero in $d$-dimensions, though, it is used to cancel 
the local UV behaviour of \Eq{eq:compact} in such a way that the locally renormalised one-loop amplitude ${\cal A}_{\r}^{(1,f)}$ 
can be obtained without altering the dimensions of the space-time
\beq
{\cal A}_{\r}^{(1,f)} = \left. {\cal A}^{(1,f)} - {\cal A}_{\uv}^{(1,f)}\right|_{d=4}~.
\label{eq:localUVoneloop}
\eeq

\section{Dual amplitude for $H\to \gamma\gamma$ at two loops}
\label{sec:hgg}

\begin{figure}[t]
	\begin{center}
		\includegraphics[width=\textwidth]{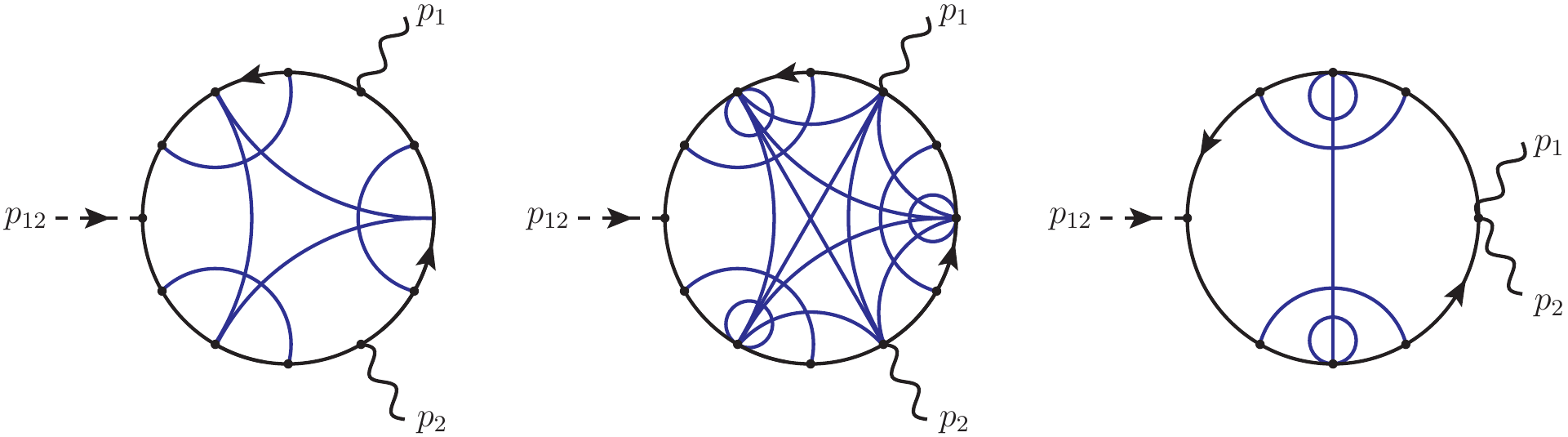}
		\caption{Two-loop mandala Feynman diagrams for $H \to\gamma\gamma$. 
			The black solid lines are quarks (left diagram) or scalars (middle and right diagrams).
			The blue solid lines are virtual photons. 
			\label{fig:mandala}}
	\end{center}
\end{figure}

At two-loop level, there are 12 Feynman diagrams contributing to the $H\to \gamma\gamma$ scattering amplitude with internal top quarks. For internal charged scalars, there are 37 Feynman diagrams. The corresponding two-loop diagrams are drawn in figure~\ref{fig:mandala} where all those that share the same global topology are superimposed in the so-called mandala diagrams. In this paper, we only consider QED corrections, and therefore photons as the extra internal particle, and do not take into account ``mixed'' diagrams where different massive particles may appear. In the traditional approach, massless snail diagrams are usually ignored, since they integrate to zero. However, within our approach, we need them to preserve the universal structure of the integrands.

All the diagrams are planar and can be constructed from the following internal momenta: 
\bea
\alpha_1: && q_i = \ell_1+p_i~, \qquad q_{12} = \ell_1 + p_{12}~,  \qquad q_3 = \ell_1~, \nn \\ 
\alpha_2: && q_4 = \ell_2~, \nn \\ 
-\alpha_2:  && q_{\overline{4}} = - \ell_2~, \nn \\ 
\alpha_3: && q_{\overline{i}} = \ell_{12} + p_i~, \qquad q_{\overline{12}} = \ell_{12} + p_{12}~, \qquad q_{\overline{3}} = \ell_{12}~.
\eea
Only $q_4$ (and $q_{\overline{4}}$) is massless (it labels the photon), while all the other internal momenta have mass $M_f$.

If the Higgs boson is on shell, the loop amplitude is below threshold and is therefore purely real.
In that kinematical regime the dual prescriptions become irrelevant, and the dual functions fulfil the identity
\beq
G_D(\alpha_i \cup \alpha_j) = G_D(\alpha_i) \, G_F(\alpha_j) + G_F(\alpha_i) \, G_D(\alpha_j)~.
\eeq
Hence, the LTD representation in \Eq{twoloopduality} adopts the simpler form
\bea
{\cal A}^{(2)}_N  &=&
\int_{\ell_1} \, \int_{\ell_2} {\cal N} (\ell_1, \ell_2,\{p_i\}_N) \, \otimes \, \bigg\{ 
  G_D(\alpha_1) \, G_D(\alpha_2) \, G_F(\alpha_3) \nn \\ &+& 
  G_F(\alpha_1) \, G_D(-\alpha_2) \, G_D(\alpha_3) +
  G_D(\alpha_1) \, G_F(\alpha_2) \, G_D(\alpha_3)  \bigg\}~. 
\label{twoloopdualitynothreshold}
\eea 

Following the algebraic reduction defined in section~\ref{sec:reduction},
\bea
{\cal A}^{(2,f)}_1 \propto \int_{\ell_1} \int_{\ell_2} \sum_{j,k} 
\left[ \frac{c^{(f)}_{a_0; a_1, \ldots, a_{5}} (p_1,p_2) }{(\kappa_j)^{a_0} (D_{i_1})^{a_1} (D_{i_2})^{a_2} \cdots (D_{i_5})^{a_{5}}} \right] \, \td{q_j, q_k}
+ {\rm perm.} 
\label{eq:algebraicreduced}
\eea
with $\kappa_j = q_{j,0}^{(+)}/M_f$,
\beq
\sum_{i=1}^{5} a_i \le 4~, \qquad a_{0} = \{2,1,0\}~.
\eeq
For a given dual or Feynman propagator, we have defined the dimensionless denominator $D_{i_l}  = (M_f^2 \, G_{F/D}(q_{i_l}))^{-1}$.
For example, in terms of these dimensionless denominators, the one-loop amplitude in \Eq{eq:compact} takes the form 
\bea
{\cal A}^{(1,f)} &=& \, g_f^{(1)} \, \int_{\ell_1} \, \Bigg[ -  \left(
\frac{\td{q_1}}{D_{12}\, D_3} + \frac{\td{q_{12}}}{D_3\, D_1} + \frac{\td{q_{3}}}{D_{12}\, D_1} \right) \, c_1^{(f)} \nn \\
&+& \left( \frac{\td{q_{12}}}{D_3} + \frac{\td{q_{3}}}{D_{12}}\right)\, \frac{c_{23}^{(f)}}{2} \Bigg] + \{p_1\leftrightarrow p_2\}~,
\label{eq:compactdi}
\eea
with
\beq
g_f^{(1)} = \frac{2}{\vev}~.
\eeq

From now on, we use a different global factor depending on whether the expressions it multiplies has been algebraically reduced or not. The usual $g_f$ will be used for unreduced expressions, and $g_f^{(L)}$ for reduced ones.

\subsection{The two-loop amplitude ${\cal A}_2^{(2,f)}$}

The two-loop amplitude $A_2^{(2,f)}$ is obtained by projecting ${\cal A}_{\mu \nu}^{(2,f)}$
using the projector $P_2^{\mu\nu}$ defined in~\Eq{Projectors}, namely $A_2^{(2,f)} = P_2^{\mu\nu} \, {\cal A}_{\mu \nu}^{(2,f)}$.
Due to gauge invariance, it has to vanish after integration. Still, it is interesting to obtain an explicit expression 
because it establishes a useful integrand relation that can be used afterwards. 
Remarkably, it can be written in a very compact form for $f=\{\phi, t\}$:
\bea
{\cal A}_2^{(2,f)}&=& g_f^{(2)} \,\int_{\ell_1} \int_{\ell_2} \,  \frac{c_3^{(f)}}{2} \, \sum_{i=1, 2, 12, 3}  \sigma(i) \Bigg[
\left( G(D_{\overline{i}}, \kappa_i, c_{4,u}^{(f)}) + F(D_{\overline{i}}, \kappa_4/\kappa_i) \right) \, \td{q_i,q_4} \nn \\ 
&-&F(-D_i,0) \, \td{q_{\overline{4}}, q_{\overline{i}}}  +
\left( G (D_4, \kappa_i, -c_{4,nu}^{(f)}) + F(D_4, -\kappa_{\overline{i}}/\kappa_i) \right) \td{q_i,q_{\overline{i}}} 
 \Bigg]~,
\label{eq:a2coefficient}
\eea
where
\begin{align}
g_f^{(2)} = 2s_{12}\,g_f^{(1)}=\frac{4s_{12}}{\vev}~, &\qquad
\sigma(i) =\begin{cases}
+1\,,~~~i\in\{1,2\}~, \\
-1\,,~~~i\in\{3,12\}~,
\end{cases} \nn \\
G(D_j,\kappa_i,c) =\frac{1}{\kappa_i^2}\left(\frac{1}{D_j}+c\right)~, &\qquad
F(D_i,r_\kappa) =\frac{1}{D_i}\left(\frac{2}{D_i}\left(1+r_\kappa\right)-1\right)~,\nn\\
\label{eq:FG}
\end{align}
and where the coefficients $c_{4,u}^{(f)}$ and$c_{4,nu}^{(f)}$ can be found in eqs.~(\ref{ctop}) and (\ref{cphi}). Notice there is a difference in mass dimensions between $g_f^{(1)}$ and $g_f^{(2)}$. This is due to the presence of the second loop measure and the additional $\tilde{\delta}$, whose product has dimension of mass squared. The following contribution 
\bea
{\cal S}_2^{(2,f)} &=& g_f^{(2)} \,\int_{\ell_1} \int_{\ell_2} \, \frac{c_3^{(f)}}{2} \, \sum_{i=1, 2, 12, 3}  \sigma(i) \, 
\frac{1}{\kappa_i^2}\,  \left( c_{4,u}^{(f)}  \, \td{q_i,q_4}  - c_{4,nu}^{(f)} \, \td{q_i,q_{\overline{i}}} \right) = 0~,
\label{eq:s2tadpoles}
\eea
vanishes irrespective of the rest of the expression for ${\cal A}_2^{(2,f)}$ because
the following subintegrals, with $n=\{0,1,2\}$, are equivalent
\beq
\int_{\ell_1} \frac{\td{q_i}}{\kappa_i^n} = \int_{\ell_1} \frac{\td{q_j}}{\kappa_j^n} =
\int_{\ell_2} \frac{\td{q_{\overline{i}}}}{\kappa_{\overline{i}}^n}=
\int_{\ell_2} \frac{\td{q_{\overline{j}}}}{\kappa_{\overline{j}}^n}~, \qquad i\ne j~.
\eeq
Consequently, the sum over integrals in \Eq{eq:s2tadpoles} vanishes.

\subsection{The two-loop amplitude ${\cal A}^{(2,f)}$}

In this section, we  apply the same transformation in \Eq{A1toA} at two-loop ($L=2$) to simplify the expressions for 
the amplitude ${\cal A}_1^{(2,f)}$. Then, and as explained in section~\ref{sec:reduction}, we perform an algebraic reduction
to express the dual representation in the form \Eq{eq:algebraicreduced} and extract the scalar coefficients $c^{(f)}_{a_0;a_1,\ldots, a_5}$.
As very few of them are indeed independent, they are simply relabeled as $c_i^{(f)}$.
We obtain very compact expressions for all the double cuts of the LTD representation.
The full expressions for the unrenormalised amplitude ${\cal A}^{(2,f)}$ are collected in appendix~\ref{ap:unrenormalised}. 
As for the one-loop case, the same expressions are valid regardless of the virtual particle circulating in the loop as a function of 
the flavour-dependent coefficients $c_i^{(f)}$.\\
\\
For the top quark and the scalar, the two independent coefficients that appeared at one loop are (see \Eq{eq:coeffoneloop})
\begin{align}
c_1^{(t)}&=-\frac{8}{d-2}+r_t~, \quad& c_{23}^{(t)}&= -\frac{4(d-4)}{d-2}~,\\
c_1^{(\phi)}&=\frac{4}{d-2}~, \quad& c_{23}^{(\phi)}&= \frac{2(d-4)}{d-2}~.
\end{align}
At two loops, they are still present, along with the following extra coefficients:
\begin{align}
\label{ctop}
&c_{4,u}^{(t)}=-\frac{d-2}{4}\,,&&c_{4,nu}^{(t)}=-\frac{d-2}{4}\,, &&c_7^{(t)}=-\frac{1}{4}(c_1^{(t)}-r_t)~, \nn \\
&c_8^{(t)}=c_1^{(t)}+\frac{(d-6)d+10}{2(d-2)}\, r_t\,,&&c_9^{(t)}=c_1^{(t)}-\frac{(d-8)d+10}{2(d-2)}\, r_t\,,&&c_{10}^{(t)}=c_1^{(t)}-\frac{(d-8)d+14}{2(d-2)}\, r_t~, \nn \\
&c_{11}^{(t)}=c_1^{(t)}+\frac{(d-8)d+18}{2(d-2)}\, r_t\,,&&c_{12}^{(t)}=-\frac{(d-4)(d-5)}{d-2}\, r_t\,,&&c_{13}^{(t)}=-\frac{(d-6)d+12}{2(d-2)}\, r_t~, \nn \\
&c_{14}^{(t)}=\frac{3}{4}\left(c_1^{(f)}-\frac{d}{3(d-2)}\, r_t \right)\,,&&c_{15}^{(t)}=-\frac{1}{2}\left(c_1^{(f)}+\frac{r_t}{2}\right)\,,&&c_{16}^{(t)}=\frac{d-8}{4}~, \nn \\
&c_{17}^{(t)}=\frac{d-4}{4}\,,&&c_{18}^{(t)}=-\frac{(d-4)^2}{4(d-2)}\,,&&c_{19}^{(t)}=\frac{1}{2}\left(c_1^{(t)}+\frac{1}{d-2}\, r_t\right)~, \nn \\
&c_{20}^{(t)}=\frac{1}{4} (c_1^{(t)}+r_t)\,,&&c_{21}^{(t)}=-\frac{2(d-4)}{d-2}+\frac{(d-10)d+18}{4(d-2)}\, r_t\,,&&c_{22}^{(t)}=-2+\frac{(d-4)d}{4(d-2)}\, r_t~,\\
\label{cphi}
&c_{4,u}^{(\phi)}=-\frac{d-2}{4}\,,&&c_{4,nu}^{(\phi)}=\frac{1}{4}\,,&&c_7^{(\phi)}=-\frac{1}{4}\, c_1^{(\phi)},\nn \\
&c_8^{(\phi)}=c_1^{(\phi)}\,,&&c_9^{(\phi)}=c_1^{(\phi)}-\frac{1}{(d-2)}\,r_\phi,&&c_{10}^{(\phi)}=c_1^{(\phi)}~,\nn \\
&c_{11}^{(\phi)}=c_1^{(\phi)}+\frac{d-4}{d-2}\, r_\phi\,,&&c_{12}^{(\phi)}=-\frac{3(d-4)}{2(d-2)}\, r_\phi\,,&&c_{13}^{(\phi)}=\frac{1}{d-2}\, r_\phi~, \nn \\
&c_{14}^{(\phi)}=\frac{3}{4}\,c_1^{(\phi)}~,&&c_{15}^{(\phi)}=-\frac{1}{2} \, c_1^{(\phi)}\,,&&c_{16}^{(\phi)}=\frac{1}{2}~,\nn \\
&c_{17}^{(\phi)}=0\,,&&c_{18}^{(\phi)}=0\,,&&c_{19}^{(\phi)}=\frac{1}{2}\, c_1^{(\phi)}~, \nn \\
&c_{20}^{(\phi)}=\frac{1}{4}\, c_1^{(\phi)}\,,&&c_{21}^{(\phi)}=-\frac{3}{d-2}\,,&&c_{22}^{(\phi)}=-\frac{1}{d-2}~.
\end{align}
Notice that these coefficients can be highly simplified in the particular case $d=4$, motivating further the study of a full four-dimensional representation of this process at two-loop level.
\section{Cancellation of integrand singularities at two loops}
\label{sec:cancellation}
In addition to be infrared safe, the scattering amplitude for $H\to \gamma\gamma$ is purely real if the Higgs boson is on shell. It is therefore completely free of soft and physical threshold singularities. Still, some of the dual propagators might go on shell inside 
the integration domain, leading to singularities of the integrand. As we demonstrated in refs.~\cite{Buchta:2014dfa,Buchta:2015wna} at one-loop level, one of the advantages of LTD is the partial cancellation of potential singularities of the integrand.  In this section, we extend the analysis of the integrand singularities of $H\to \gamma\gamma$ at two-loop level.

If a dual propagator $G_D(q_i;q_j)$ becomes on shell where both internal momenta $q_i$ and $q_j$  belong to the same loop line $\alpha_k$,
then the corresponding singular behaviour is equivalent to the one-loop case. For example, consider the dual propagator
$G_D(q_{\overline{12}};q_{\overline{3}})$.  A physical threshold (forward-backward singularity\footnote{A forward-backward singularity
in the loop momentum space arises in the intersection of the forward on-shell hyperboloid or positive energy mode of one propagator with 
the backward on-shell hyperboloid or negative energy mode of another propagator.}) occurs if (see ref.~\cite{Buchta:2014dfa})
\beq
k_{ji}^2 - (m_j+m_i)^2 \ge 0~, \qquad k_{ji,0} < 0~,
\label{eq:fbsingularity}
\eeq
with $k_{ji} = q_{\overline{3}}-q_{\overline{12}} = - p_{12}$ and $m_j=m_i=M_f$. Therefore, $- p_{12,0}<0$ but $s_{12}-4M_f^2 < 0$
for $s_{12} = M_H^2$. An integrand singularity (forward-forward singularity) occurs if 
\beq
k_{ji}^2 - (m_j-m_i)^2 \le 0~.
\eeq
This would be the case of, e.g., $G_D(q_{\overline{1}}; q_{\overline{12}})$ because $k_{ji}=p_2$, but this integrand 
singularity cancels in the sum of the two dual contributions involving $\td{q_{\overline{1}}}$ and $\td{q_{\overline{12}}}$.
The analysis can be extended easily to the other dual cuts, and similar conclusions are found.

Now, we consider the genuine two-loop case where a dual propagator becomes on shell in the double cut of two propagators that do not belong to the same loop line. For the rest of this section, we consider
\beq
q_i\in\alpha_1\,,\quad q_j\in\alpha_2\,\quad\text{and}\quad q_k\in\alpha_3~.
\eeq
We study study the quantity\footnote{A rigorous analysis would require to consider either $G_D(q_i;q_k)$ or $G_D(q_j;q_k)$.
For the sake of simplicity, $G_D(q_i,q_j; q_k)$ denotes the dual propagator obtained by ignoring 
the imaginary dual prescription, with both $q_i$ and $q_j$ on shell. This simplification is valid because 
we are dealing with a real scattering amplitude.}
\begin{equation}
S_{ijk}=\left\{\frac{1}{2\pi i}\tilde{\delta}(q_i,q_j)\, G_D(q_i,q_j;q_k)\right\}+\{\overline{j},k,i\}+\{k,i,j\}\,,
\end{equation}
where $\overline{j}$ indicates that we reverse the momentum flow of $q_j$, as explained in section~\ref{sec:twoloops}, namely
\beq
\label{ReverseFlowExample}
q_{\overline{j}}=-q_j~, \qquad \Su{q_{\overline{j},0}}=\Su{q_{j,0}}~.
\eeq
We therefore have
\begin{align}
S_{ijk}
&=\frac{\delta(q_{i,0}-\Su{q_{i,0}})}{2\Su{q_{i,0}}}\frac{\delta(q_{j,0}-\Su{q_{j,0}})}{2\Su{q_{j,0}}}\frac{1}{(\Su{q_{i,0}}+\Su{q_{j,0}}+k_{k(ij),0})^2-(\Su{q_{k,0}})^2}\nn\\
&+\frac{\delta(q_{j,0}+\Su{q_{j,0}})}{2\Su{q_{j,0}}}\frac{\delta(q_{k,0}-\Su{q_{k,0}})}{2\Su{q_{k,0}}}\frac{1}{(\Su{q_{j,0}}+\Su{q_{k,0}}-k_{k(ij),0})^2-(\Su{q_{i,0}})^2}\nn\\
&+\frac{\delta(q_{i,0}-\Su{q_{i,0}})}{2\Su{q_{i,0}}}\frac{\delta(q_{k,0}-\Su{q_{k,0}})}{2\Su{q_{k,0}}}\frac{1}{(\Su{q_{k,0}}-\Su{q_{i,0}}-k_{k(ij),0})^2-(\Su{q_{j,0}})^2}\,,\label{SijkValue}
\end{align}
where
\beq
\label{kk(ij)}
k_{k(ij)}=q_k-q_i-q_j
\eeq
only depends on external momenta. The quantity $S_{ijk}$ becomes singular in the limits summarised by the condition $\lambda_{\pm\pm\pm} \to 0$, with 
\beq
\lambda_{\pm\pm\pm}= \pm \Su{q_{i,0}} \pm \Su{q_{j,0}} \pm \Su{q_{k,0}} + k_{k(ij),0}~.
\eeq
Accorded to \Eq{SijkValue}, only four independent solutions have to be considered, namely $\lambda_{+++}$, $\lambda_{++-}$, $\lambda_{+--}$ and $\lambda_{---}$.
For two of these limits, there is a perfect cancellation of the integrand singularities, as indeed
\beq
\lim_{\lambda_{++-}\to 0} S_{ijk} = {\cal O} (\lambda^0_{++-})~, \qquad \lim_{\lambda_{+--} \to 0} S_{ijk} = {\cal O} (\lambda^0_{+--})~,
\eeq
while for the remaining two limits,
\bea
&& \lim_{\lambda_{+++}\to 0} S_{ijk} = - \theta(-k_{k(ij),0}) \, \lambda^{-1}_{+++} \, 
\frac{\delta(q_{i,0} - \Su{q_{i,0}}) \, \delta(q_{j,0} - \Su{q_{j,0}})}{(2\Su{q_{i,0}})(2\Su{q_{j,0}})(2\Su{q_{k,0}})} + {\cal O} (\lambda^0_{+++})~, \nn \\
&& \lim_{\lambda_{---}\to 0} S_{ijk} = \theta(k_{k(ij),0}) \, \lambda^{-1}_{---} \, 
\frac{\delta(q_{j,0} + \Su{q_{j,0}}) \, \delta(q_{k,0} - \Su{q_{k,0}})}{(2\Su{q_{i,0}})(2\Su{q_{j,0}})(2\Su{q_{k,0}})} + {\cal O} (\lambda^0_{---})~.
\eea
Although these singularities remain in the general case, it is possible to show that $\lambda_{+++}$ (resp. $\lambda_{---}$) can only cancel if, in the particular case where $m_i=m_k=M_f$ and $m_j=0$,
\begin{equation}\label{LambdaCond}
\begin{cases}
k_0<0\\
k^2-4M_f^2\geq0
\end{cases}
\qquad\left(\text{resp.}
\quad
\begin{cases}
k_0>0\\
k^2-4M_f^2\geq0
\end{cases}
\right)~.
\end{equation}
Because of the fact $k$ does not depend on the loop momenta, the highest possible value of $k^2$ is $(p_1+p_2)^2=s_{12}$, reached for instance when considering $G_D(q_1,q_4;q_{\overline{12}})$. This means the second condition of \Eq{LambdaCond} reduces to
\begin{equation}
\frac{4M_f^2}{s_{12}}<1~,
\end{equation}
which is not fulfilled if, as stated above, the Higgs boson is assumed to be on shell, i.e. $s_{12}=M_H^2$.\\

Now, we consider the possibility to encounter soft singularities as one of the internal particles is massless. 
The integrand becomes soft in the loop momentum $\ell_2$ if $q_{j,0}^{(+)} = 0$. In that case, the analysis of the singular behaviour is
very similar to the one-loop case. We should solve the condition
\beq
\lambda_{\pm\pm}= \pm \Su{q_{i,0}} \pm \Su{q_{k,0}} + k_{k(ij),0}
\label{eq:softtwoloop}
\eeq
 in $\ell_1$. If the three propagators are attached to the same vertex, then  $k_{k(ij),0}=0$
and $\Su{q_{i,0}} = \Su{q_{k,0}}$. In that case $\lambda_{+-}$ and $\lambda_{-+}$ can vanish, 
then enhancing the integrand singularity in $\ell_2$, but the overall singularity cancels between dual contributions. 
The other solution, with $\lambda_{++} =0$, is not possible 
because $\Su{q_{i,0}} = \Su{q_{k,0}} \ge M_f$. If the three propagators do not interact in the same vertex, 
then $k_{k(ij),0}\ne 0$. This configuration includes the cases where $q_i$ and $q_k$ belong to the same or to 
different loop lines. There are solutions to \Eq{eq:softtwoloop}, but again either they cancel among dual contributions
or \Eq{eq:fbsingularity} is not fulfilled. In all the cases, the soft singularities of the integrand 
in $\ell_2$ do not translate into soft singularities of the amplitude because of the integration measure.

\section{Algorithmic approach to two-loop local UV renormalisation}
\label{sec:UVren}
Let us consider a generic unrenormalised two-loop amplitude ${\cal A}^{(2)}$, written as
\beq
{\cal A}^{(2)}=\int_{\ell_ 1}\,\int_{\ell_2}\,\mathcal{I}(\ell_1,\ell_2)~,
\eeq
which we will assume to be completely free of any infrared singularity. Thus, only UV singularities may appear when either or both of 
$|\boldsymbol{\ell}_1|$ and $|\boldsymbol{\ell}_2|$ go to infinity. The local two-loop UV counter-terms are built recursively by first
fixing one of the two loop momenta, say $\ell_j$, and expanding the integrand $\mathcal{I}(\ell_1,\ell_2)$ up to logarithmic order 
around the UV propagator~\cite{Becker:2010ng}
\begin{equation}
G_F(q_{i,\uv})=\frac{1}{q_{i,\uv}^2-\mu_\uv^2 + \ii}~,
\end{equation}
where the arbitrary scale $\mu_\uv$ represents the renormalisation scale, and 
$q_{i,\uv} = \ell_i+k_{i,\uv}$. For simplicity, we take $k_{i,\uv}=0$. The quantity
\beq
\label{ACT1CT2}
{\cal A}^{(2)}-{\cal A}^{(2)}_{1,\uv}-{\cal A}^{(2)}_{2,\uv}~,
\eeq
where ${\cal A}^{(2)}_{i,\uv}$ denotes the two-loop amplitude ${\cal A}^{(2)}$ in the limit 
$|\boldsymbol{\ell}_i| \to \infty$, is not necessarily UV finite when both loop momenta are simultaneously large. It is necessary to 
subtract also the double UV behaviour  of \Eq{ACT1CT2}. With this contribution, which is represented by ${\cal A}^{(2)}_{\uv^2}$, 
the final renormalised amplitude reads 
\beq
\label{ACT1CT2CT112}
{\cal A}^{(2)}_{\r}= {\cal A}^{(2)}-{\cal A}^{(2)}_{1,{\rm UV}}-{\cal A}^{(2)}_{2,{\rm UV}}-{\cal A}^{(2)}_{{\rm UV}^2}~,
\eeq
and is UV safe in all the limits.  As an example, we consider
\beq
{\cal I}(\ell_1,\ell_2)=\frac{1}{(\ell_1^2-M^2+\ii)(\ell_2^2-M^2+\ii)^2((\ell_1+\ell_2)^2-M^2+\ii)}~.
\eeq
This integrand produces a UV singularity when $|\boldsymbol{\ell_1}|\to\infty$, but it is superficially regular in $\ell_2$, meaning ${\cal A}_{2,\uv}^{(2)} =0$. Computing the remaining counter-term gives
\beq
{\cal A}_{1,\uv}^{(2)}=\int_{\ell_ 1}\, \frac{1}{(\ell_{1}^2-\mu_\uv^2+\ii)^2} \, 
\,\int_{\ell_2}\,\frac{1}{(\ell_2^2-M^2+\ii)^2}~,
\eeq
which effectively removes the UV behaviour in $\ell_1$, but at the same time also introduces a singularity in $\ell_2$. 
It is therefore necessary to introduce the additional counter-term ${\cal A}_{{\rm UV}^2}^{(2)}$ to fix the UV behaviour when 
both loop momenta go to infinity. This is done by expanding \Eq{ACT1CT2} for very high values of $\ell_1$ and $\ell_2$, 
while never neglecting one compared to the other. In this example, we get
\begin{equation}
{\cal A}_{\uv^2}^{(2)}=\int_{\ell_ 1}\,\int_{\ell_2}\,\frac{1}{(\ell_1^2-\mu_\uv^2+\ii)(\ell_2^2-\mu_\uv^2+\ii)^2}
\left(\frac{1}{(\ell_1+\ell_2)^2-\mu_\uv^2+\ii}-\frac{1}{\ell_1^2-\mu_\uv^2+\ii}\right)~.
\end{equation}
Then, the renormalised amplitude, as defined in \Eq{ACT1CT2CT112}, is finite in the UV. It is still necessary, though, to introduce subleading contributions to fix the renormalisation scheme. This is better explained in the following for the $H\to \gamma \gamma$ two-loop amplitude.

With the labelling of the internal momenta that we have adopted for the $H\to \gamma \gamma$ amplitude, 
it is more convenient to express the UV behaviour at two loops in terms of $q_{1,\uv} = \ell_1$ and $q_{12,\uv} = \ell_{12}$, 
with $\ell_2=\ell_{12}-\ell_1$.
Explicitly, the single and double UV behaviours are implemented by making use of the following transformations
\bea\label{TransfUV}
{\cal S}_{j,\uv} &:& \{ \ell_j^2 ~|~  \ell_j\cdot k_i \}  \to \{ 
\lambda^2 \, q_{j,\uv}^2 + (1-\lambda^2) \, \mu_{\uv}^2 ~|~ \lambda \, q_{j,\uv}\cdot k_i \}~,  \qquad j,k \in \{1, 12\}~, \nn\\ 
{\cal S}_{\uv^2} &:& \{ \ell_j^2 ~|~ \ell_j \cdot \ell_k ~|~ \ell_j\cdot k_i \}  \to  \nn \\ 
&& \{ \lambda^2 \, q_{j, \uv}^2 + (1-\lambda^2) \, \mu_{\uv}^2 ~|~
\lambda^2 \, q_{j,\uv}\cdot q_{k,\uv} + (1-\lambda^2)/2 \, \mu_{\uv}^2 ~|~
\lambda \, q_{j, \uv}\cdot k_i \}~,  
\eea
then expanding for $\lambda\to\infty$ and truncating the corresponding series in $\lambda$ up to logarithmic degree. 
This last operation is represented by the function $L_\lambda$.
In particular, the UV counter-terms are defined as
\bea
{\cal A}^{(2,f)}_{j,\uv}&=& L_\lambda\left(\left. {\cal A}^{(2,f)}\right|_{{\cal S}_{j,\uv}}\right) - 
(e\, e_f)^2\left( d_{j,\uv}^{(f)} \, \mu_{\uv}^2 \, \int_{\ell_j} \left(G_F(q_{j,\uv}) \right)^3 \right) {\cal A}^{(1,f)}~, \quad j\in \{1, 12\}~,\\
{\cal A}^{(2,f)}_{\uv^2} &=& L_\lambda\left(\left. \left({\cal A}^{(2,f)}- \sum_{j=1,12} {\cal A}^{(2,f)}_{j,\uv}\right)\right|_{{\cal S}_{\uv^2}}\right)\nn\\
&-& 4 g_f\,s_{12}\,(e\, e_f)^2 \left(d_{\uv^2}^{(f)} \, \mu_{\uv}^4\int_{\ell_1}\int_{\ell_2} \left(G_F(q_{1,\uv}) \right)^3\left(G_F(q_{12,\uv}) \right)^3\right)~,\label{AUV2}
\eea
where ${\cal A}^{(1,f)}$ is the unintegrated one-loop amplitude written in terms of $\ell_i$ ($\ell_1$ or $\ell_{12}$) with $i\ne j$, and where $d_{j,\uv}^{(f)}$ and $d_{\uv^2}^{(f)}$ are scalar coefficients used to fix the renormalisation scheme. Note that the integrals they multiply integrate to finite quantities. The factor 4 appearing in the second line of \Eq{AUV2} is arbitrary and has been introduced to conveniently rescale $d_{\uv^2}^{(f)}$. 

\begin{table}
\begin{center}
\begin{tabular}{|c|ccc|ccc|cc|cc}
\hline
& $c_{H,\uv}^{(f)}$ & $d_{H, \uv}^{(f)}$ & $C_{H, \uv}^{(f)}$ 
& $c_{\gamma, \uv}^{(f)}$& $d_{\gamma, \uv}^{(f)}$ & $C_{\gamma, \uv}^{(f)}$ \\ \hline \hline
$t \bar{t}$ & $d$ & $4$ & $4$ & $(d-2)/2$ & $2$ & $1$  \\
$\phi\phi^\dagger$ & 1 & $0$ & $1$ & $-2~~~	$ & $0$ & $-2~~~$ \\

\hline
\end{tabular}
\caption{Values of the scheme fixing paremeters in the $\overline{\text{MS}}$ for the single UV counter-terms.
\label{tab:renorm}}
\end{center}
\end{table}

\subsection{Higgs boson vertex renormalisation}
\label{sec:higgsvertex}

In the Feynman gauge, the one-loop QED correction to the Higgs boson vertex exhibits the UV behaviour 
\bea
{\bf \Gamma}^{(1,f)}_{H, \uv} &=& (e \, e_f)^2 \, \int_{\ell_1} \left( G_F(q_{1,\uv})\right)^2 \, 
\left(c_{H,\uv}^{(f)}  - G_F(q_{1,\uv}) \, d_{H, \uv}^{(f)} \,  \mu_{\uv}^2\right)  \, {\bf \Gamma}^{(0,f)}_{H} \nn \\
&=& (e \, e_f)^2  \frac{\Se}{16\pi^2} \left( \frac{\mu_\uv^2}{\mu^2}\right)^{-\ep} \frac{C_{H,\uv}^{(f)}}{\ep} \, {\bf \Gamma}^{(0,f)}_{H}~,
\label{eq:genericvertex}
\eea 
where ${\bf \Gamma}^{(0,t)}_{H} = - \imath \, M_t/\vev$ and ${\bf \Gamma}^{(0,\phi)}_{H} = -2 \imath \, M_\phi^2/\vev$ are the tree-level vertex interactions, and $\Se=(4\pi)^\epsilon\Gamma(1+\epsilon)$.
The coefficients $d_{H,\uv}^{(f)}$ are subleading and are necessary to fix the  renormalisation scheme. 
The values of these coefficients in the $\overline{\text{MS}}$ renormalisation scheme\footnote{We distinguish $\Se= (4\pi)^\ep \, \Gamma(1+\ep)$ from the usual $\overline{\text{MS}}$ scheme factor $S_\ep^{\overline{\text{MS}}} = (4\pi)^\ep \exp(-\ep \gamma_E)$ or $S_\ep= (4\pi)^\ep/\Gamma(1-\ep)$ as used in ref.~\cite{Bolzoni:2010bt}. At NLO all these definitions lead to the same expressions. At NNLO, they lead to slightly different bookkeeping of the IR and UV poles at intermediate steps, but physical cross-sections 
of infrared safe observables are the same.} are summarised in table~\ref{tab:renorm}, and they are related
to the coefficient of the integrated vertex counter-term through
\beq
C_{H,\uv}^{(f)} = c_{H,\uv}^{(f)}+\frac{\ep}{2} \, d_{H, \uv}^{(f)}~.
\label{eq:integratedHvertex}
\eeq
From the expression of the vertex counter-term in \Eq{eq:genericvertex} we can construct the UV counter-term of the two-loop scattering amplitude in the limit $| \boldsymbol{\ell}_1| \to \infty$ with $|\boldsymbol{\ell}_{12}|$ fixed. It reads
\beq
{\cal A}_{H,\uv}^{(2,f)} = \int_{\ell_1}  
\left( G_F(q_{1,\uv})\right)^2 \, \left(c_{H,\uv}^{(f)} - G_F(q_{1,\uv}) \, d_{H,\uv}^{(f)} \, \mu_{\uv}^2\right)
\, {\cal A}^{(1,f)}(\ell_{12})~,
\label{A1UV}
\eeq
where ${\cal A}^{(1,f)}$ is the unrenormalised one-loop $H\to\gamma\gamma$ amplitude in \Eq{eq:compact}. Note that ${\cal A}^{(1,f)}$ is locally divergent and should be renormalised as well in \Eq{A1UV}. However, by definition, we want ${\cal A}_{H,\uv}^{(2,f)}$ to exactly cancel the singularities arising when  $|\boldsymbol{\ell_1}|\to\infty$. Putting ${\cal A}_{\r}^{(1,f)}$ instead of ${\cal A}^{(1,f)}$ in \Eq{A1UV} would therefore alter the UV behaviour of the single counter-term and not properly remove the corresponding infinities. It is only when considering the double UV counter-term (section~\ref{sec:doubleUV}) that the one-loop amplitude implicitly gets renormalised.

The corresponding dual representation is
\beq
{\cal A}_{H,\uv}^{(2,f)} (q_{1,\uv},q_{\overline{i}}) =
\int_{\ell_1}  \frac{\td{q_{1,\uv}}}{2\, (q_{1,\uv}^{(+)})^2} \, \left(c_{H,\uv}^{(f)} + d_{H,\uv}^{(f)} \, \frac{3 \mu_{\uv}^2}{4\, (q_{1,\uv}^{(+)})^2} \right)
\, {\cal A}^{(1,f)}(q_{\overline{i}})~,
\eeq
where $q_{1,\uv}^{(+)} = \sqrt{\boldsymbol{\ell}_1^2+\mu_{\uv}^2}$. Since the diagrams (2 for the top quark, 3 for the charged scalar) that contribute to the Higgs vertex correction are the only ones that are divergent when $|\boldsymbol{\ell_1}|\to\infty$, we directly have ${\cal A}_{1,\uv}^{(2,f)} = {\cal A}_{H,\uv}^{(2,f)}$.
\\
\\

\subsection{Photon vertex renormalisation}

The one-loop correction to the photon interaction vertex to top quarks in the UV is given in the Feynman gauge by
\bea
{\bf \Gamma}^{(1,t)}_{\gamma, \uv} 
&=& (e\, e_t)^2 \, \int_{\ell_2} \left( G_F(q_{12,\uv})\right)^2 \, 
\left( \left( c_{\gamma, \uv}^{(t)} - G_F(q_{12,\uv}) \, d_{\gamma, \uv}^{(t)} \, \mu_{\uv}^2 \right) \, {\bf \Gamma}^{(0,t)}_{\gamma} 
+ c_{\gamma, \uv}^{(t)}  \, {\bf \Delta}^{(1,t)}_{\gamma, \uv}\right) \nn \\
&=& (e \, e_t)^2 \frac{\Se}{16\pi^2} \left( \frac{\mu_\uv^2}{\mu^2}\right)^{-\ep} \frac{C_{\gamma,\uv}^{(t)}}{\ep} \, {\bf \Gamma}^{(0,t)}_{\gamma}~,
\label{eq:ttphoton}
\eea
with
\beq
{\bf \Delta}^{(1,t)}_{\gamma, \uv} = 
{\bf \Gamma}^{(0,t)}_{\gamma} - 2 \, G_F(q_{12,\uv}) \, \bq_{12,\uv} \, \{ \bq_{12,\uv}, {\bf \Gamma}^{(0,t)}_{\gamma}\}~.
\label{eq:ttphotondelta}
\eeq
In \Eq{eq:ttphoton}, the term proportional to ${\bf \Delta}^{(1,t)}_{\gamma, \uv}$ integrates to zero
in $d$ space-time dimensions.
Similarly to \Eq{eq:integratedHvertex}, the coefficient of the integrated vertex counter-term is given by 
\beq
C_{\gamma,\uv}^{(f)} = c_{\gamma,\uv}^{(f)}+ \frac{\ep}{2} \, d_{\gamma, \uv}^{(f)}~.
\eeq
Although the integrated UV vertex correction is proportional to the tree-level vertex 
${\bf \Gamma}^{(0,t)}_{\gamma} = \imath \, e \, e_t \, \gamma^{\mu_i}$,
thanks to the replacement $q_{12,\uv}^{\mu_1}\, q_{12,\uv}^{\mu_2} \to q_{12,\uv}^2 \, g^{\mu_1\mu_2}/d$, we cannot use this replacement in the unintegrated form because it would alter the local UV behaviour. We must keep the full expression in \Eq{eq:ttphoton}, including especially the term proportional to ${\bf \Delta}^{(1,t)}_{\gamma, \uv}$, to construct the local UV counter-term of the two-loop scattering amplitude.

For charged scalars as internal particles, we need to consider both the three-point and the four-point interaction vertices. For the three-point vertex there are three contributing diagrams, and the corresponding counter-term reads
\bea
{\bf \Gamma}^{(1,\phi)}_{\gamma, \uv} &=& (e\, e_\phi)^2 \, 
\int_{\ell_2} \left( G_F(q_{12, \uv})\right)^2 \, 
c_{\gamma, \uv}^{(\phi)} \, \left( {\bf \Gamma}^{(0,\phi)}_{\gamma} +  {\bf \Delta}^{(1,\phi)}_{\gamma, \uv} \right) \nn \\
&=& (e \, e_\phi)^2\frac{\Se}{16\pi^2}
\left( \frac{\mu_\uv^2}{\mu^2}\right)^{-\ep} \frac{C_{\gamma,\uv}^{(\phi)}}{\ep} \, {\bf \Gamma}^{(0,\phi)}_{\gamma}~,
\label{eq:scalarphoton}
\eea
where
\beq
{\bf \Delta}^{(1,\phi)}_{\gamma, \uv} = \frac{1}{2} \,
\left( {\bf \Gamma}^{(0,\phi)}_{\gamma} (q_{12,\uv},p_i) + 4 \, \imath (e\,  e_\phi) \, G_F(q_{12,\uv}) \, (q_{12,\uv}\cdot (k_{i-1}+ k_i) )  \,  q_{12,\uv}^{\mu_i} \right)~,
\label{eq:scalarphotondelta}
\eeq
with ${\bf \Gamma}^{(0,\phi)}_{\gamma}= - \imath \, (e\,  e_\phi) \left( q_{i-1}+ q_i \right)^{\mu_i}$
and ${\bf \Gamma}^{(0,\phi)}_{\gamma} (q_{12,\uv},p_i) = - \imath \, (e\,  e_\phi) \left( q_{\overline{i-1}}+ q_{\overline{i}} \right)^{\mu_i}$, where $q_{i-1}$ and $q_i$ are the outgoing and incoming internal momenta, respectively. For example, $q_{i-1}=q_3+p_1$ and $q_i=q_3+p_{12}$ for the vertex correction with emission of a photon with momentum $p_2$ in the lower corner of the two-loop Feynman diagram.

For the four-point interaction vertex, there are nine contributing diagrams\footnote{Note that these diagrams include the ones contributing to the Higgs vertex correction. However, the singular regime considered here is different since we study the limit $|\boldsymbol{\ell_{12}}|\to\infty$.}, and we have
\bea
{\bf \Gamma}^{(1,\phi)}_{\gamma \gamma, \uv} &=& (e\, e_\phi)^2 \,  
\int_{\ell_2} \left( G_F(q_{12, \uv})\right)^2\, c_{\gamma, \uv}^{(\phi)} \, 
\left( {\bf \Gamma}^{(0,\phi)}_{\gamma\gamma}  + {\bf \Delta}^{(1,\phi)}_{\gamma \gamma, \uv} \right)\nn \\
&=& (e \, e_\phi)^2 \frac{\Se}{16\pi^2} 
\left( \frac{\mu_\uv^2}{\mu^2}\right)^{-\ep} \frac{C_{\gamma,\uv}^{(\phi)}}{\ep} \, {\bf \Gamma}^{(0,\phi)}_{\gamma\gamma}~,
\label{eq:scalarphotonphoton}
\eea
where
\beq
{\bf \Delta}^{(1,\phi)}_{\gamma \gamma, \uv} =  \frac{1}{2} \, {\bf \Gamma}^{(0,\phi)}_{\gamma\gamma} 
- 4\, \imath \, (e\, e_\phi)^2 \, G_F(q_{12, \uv}) \, q_{12, \uv}^{\mu_1} \, q_{12, \uv}^{\mu_2}~.
\label{eq:scalarphotonphotondelta}
\eeq
with  ${\bf \Gamma}^{(0,\phi)}_{\gamma\gamma} = 2\, \imath \, (e\, e_\phi)^2 \,  g^{\mu_1\mu_2}$. 
Remarkably, the coefficient $c_{\gamma, \uv}^{(\phi)}$ is the same as for the three-point interaction vertex. 
Again, in \Eq{eq:scalarphoton} and \Eq{eq:scalarphotonphoton}, we cannot apply the replacement $q_{12,\uv}^{\mu_1}\, q_{12,\uv}^{\mu_2} \to q_{12,\uv}^2 \, g^{\mu_1\mu_2}/d$ at integrand level even though the terms ${\bf \Delta}^{(1,\phi)}_{\gamma}$ in \Eq{eq:scalarphoton}  and  ${\bf \Delta}^{(1,\phi)}_{\gamma\gamma}$ in \Eq{eq:scalarphotonphoton} integrate to zero.
Also notice that it was not necessary to introduce subleading terms for the scalar vertices because the finite part of the corresponding integrated counter-term is already 0.

The integrated counter-term reads
\bea
{\cal A}_{\gamma,\uv}^{(2,f)} (q_i, q_{12,\uv}) &=&\left( \frac{\Se}{16\pi^2} \left( \frac{\mu_\uv^2}{\mu^2}\right)^{-\ep} \frac{2C_{\gamma,\uv}^{(f)}}{\ep} \right){\cal A}_\gamma^{(1,f)}(q_i)~,\label{AgammaUV}\\
{\cal A}_{\gamma\gamma,\uv}^{(2,\phi)} (q_i, q_{12,\uv}) &=& \left( \frac{\Se}{16\pi^2} \left( \frac{\mu_\uv^2}{\mu^2}\right)^{-\ep} \frac{C_{\gamma,\uv}^{(\phi)}}{\ep} \right){\cal A}_{\gamma\gamma}^{(1,\phi)}(q_i)~, \label{AgammagammaUV}
\eea
where ${\cal A}_{\gamma}^{(1,f)}$ is the sum of the two one-loop amplitudes involving triangle diagrams and ${\cal A}_{\gamma\gamma}^{(1,\phi)}$ is the one-loop bubble amplitude (which appears only for the charged scalar). The relative factors 2 in \Eq{AgammaUV} comes from the fact there are two three-point vertices to renormalise for each contributing diagram.

The corresponding dual representations are
\bea
{\cal A}_{\gamma,\uv}^{(2,f)} (q_i,q_{12,\uv}) &=&
\int_{\ell_1}  \frac{\td{q_{12,\uv}}}{(q_{12,\uv}^{(+)})^2} \, \left[ \left(c_{\gamma,\uv}^{(f)} + d_{\gamma,\uv}^{(f)} \, \frac{3 \mu_{\uv}^2}{4\, (q_{12,\uv}^{(+)})^2} \right)
\, {\cal A}_\gamma^{(1,f)}(q_i) + c_{\gamma,\uv}^{(f)} \,  {\bf \Delta}^{(1,f)}_{\gamma, \uv}(q_i,q_{12,\uv}) \right]~,\nn\\
\label{AgammaUVDual}\\
{\cal A}_{\gamma\gamma,\uv}^{(2,\phi)} (q_i,q_{12,\uv}) &=&
\, \int_{\ell_1}  \frac{\td{q_{12,\uv}}}{2\, (q_{12,\uv}^{(+)})^2} \, c_{\gamma,\uv}^{(\phi)} \left[
{\cal A}_{\gamma\gamma}^{(1,\phi)}(q_i) + \,  {\bf \Delta}^{(1,\phi)}_{\gamma\gamma, \uv}(q_i,q_{12,\uv}) \right]~,\label{AgammagammaUVDual}
\eea
where $q_{12,\uv}^{(+)} = \sqrt{\boldsymbol{\ell}_{12}^2+\mu_{\uv}^2}$. 

\subsection{Self-energies renormalisation}

With our labelling of the momenta the self-energy insertions are defined in terms of the 
internal momenta $q_4=\ell_2$ and $q_{\overline{i}}= \ell_{12} + k_i$, with $i=\{1,2,12,3\}$.
Explicitly, in the Feynman gauge we have (notice the relative sign in $q_{\overline{i}}$ with respect to \cite{Sborlini:2016hat} because of the fact the momentum flows in the opposite direction)
\beq
\Sigma^{(1,t)}(q_i) = (e\, e_t)^2  \int_{\ell_2} G_F(q_4, q_{\overline{i}}) \,  \left( - 2 \, c_{\gamma,\uv}^{(t)} \, \bq_{\overline{i}} + c_{H,\uv}^{(t)}  \, M_t \right)~.
\label{selfqi}
\eeq
The UV expansion of \Eq{selfqi} reads
\bea
\Sigma_{\uv}^{(1,t)}(q_i) &=& (e\, e_t)^2  \int_{\ell_2} ( G_F(q_{12,\uv}) )^2 \,  \bigg[ -  
\left( c_{\gamma,\uv}^{(t)} - G_F(q_{12,\uv}) \, d_{\gamma,\uv}^{(t)} \, \mu_{\uv}^2 \right) \, \bq_{i}  \nn \\ 
&+& \left( c_{H,\uv}^{(t)} - G_F(q_{12,\uv}) \, d_{H,\uv}^{(t)} \, \mu_{\uv}^2 \right) \, M_t   
+ c_{\gamma,\uv}^{(t)} \,  {\bf \Delta}^{(1,t)}_{\Sigma, \uv} \bigg] \nn \\
&=&  (e \, e_t)^2\frac{\Se}{16\pi^2} \left( \frac{\mu_{\uv}^2}{\mu^2} \right)^{-\ep} \frac{1}{\ep} \left( - C_{\gamma,\uv}^{(t)} \, \bq_i + C_{H,\uv}^{(t)}  \, M_t\right)~,  
\label{newselfUV}
\eea
with 
\beq
{\bf \Delta}^{(1,t)}_{\Sigma, \uv} =  2\left( \frac{\bq_i}{2} - \bk_i  - \bq_{12,\uv}  
- 4 \, G_F(q_{12,\uv}) \left(q_{12,\uv} \cdot(\frac{q_i}{2}-k_i)\right) \, \bq_{12,\uv} \right)~,  
\label{newselfUVdelta}
\eeq
where the coefficients $d_{k,\uv}^{(t)}$ are subleading contributions to be fixed through the renormalisation scheme.
It is remarkable that it has been possible to write the quark self-energy in terms of the same coefficients that appear in the Higgs boson and photon vertices. Notice that the expression in \Eq{newselfUV} is simpler than the corresponding expression provided in ref.\cite{Sborlini:2016hat}
and only differs at ${\cal O} (\ep)$, which does not have any consequence at the considered order.

The scalar self-energy corrections, which also include the snail diagrams, is written
\beq
\Sigma^{(1,\phi)}(q_i)= (e\, e_\phi)^2  \int_{\ell_2} G_F(q_4) \left( - c_{T,\uv}^{(\phi)}+ G_F(q_{\overline{i}}) \,  \left(  -  c_{\gamma,\uv}^{(\phi)} \, 
\left(q_{\overline{i}} + \frac{q_i}{2}\right) \cdot q_i  + c_{H,\uv}^{(\phi)}\, M_\phi^2 \right) \right)~,  
\label{selfqiscalar}
\eeq
where $c_{T,\uv}^{(\phi)} = d-1$, or, equivalently, $c_{T,\uv}^{(\phi)} = 4 (c_{4,nu}^{(\phi)} - c_{4,u}^{(\phi)})$.  
One possibility would be to subtract the contribution which is proportional to $c_{T,\uv}^{(\phi)}$ before expanding in the UV, which would be equivalent to subtract a zero (actually this would not only work for the term generated by the snail diagrams, but also for any term that exclusively contains propagators depending only on $\ell_1$). However, it would modify only the double cuts $\td{q_i,q_4}$.
The UV expansion of \Eq{selfqiscalar} reads 
\bea
\Sigma_{\uv}^{(1, \phi)}(q_i) &=& (e\, e_\phi)^2  \int_{\ell_2} G_F(q_{12,\uv})  \,  \bigg[ 
G_F(q_{12,\uv})  \, \left( - c_{\gamma,\uv}^{(\phi)} \, q_i^2 + c_{H,\uv}^{(\phi)} \, M_\phi^2 
+ c_{\gamma,\uv}^{(\phi)} \,  {\bf \Delta}^{(1,\phi)}_{\Sigma, \uv}  \right)
+ c_{T,\uv}^{(\phi)} \,  {\bf \Delta}^{(1,\phi)}_{T, \uv}  \bigg]\nn \\
&=&  (e \, e_\phi)^2 \frac{\Se}{16\pi^2} \, \left( \frac{\mu_{\uv}^2}{\mu^2} \right)^{-\ep} \frac{1}{\ep} \left( - C_{\gamma,\uv}^{(\phi)}\, q_i^2+ C_{H,\uv}^{(\phi)} M_\phi^2\right)~, 
\label{newselfUVscalar}
\eea
with
\beq
{\bf \Delta}^{(1,\phi)}_{\Sigma, \uv} =   \left( \frac{q_i}{2}- k_i - q_{12,\uv} \right) \cdot q_i 
- 4 \, G_F(q_{12,\uv}) \, \left(q_{12,\uv} \cdot \left(\frac{q_i}{2} - k_i\right) \right) (q_{12,\uv} \cdot q_i)~, 
\label{newselfUVscalardelta}
\eeq
\bea
{\bf \Delta}^{(1,\phi)}_{T, \uv} &=&  - 2 + G_F(q_{12,\uv}) \,  \left(  (q_{12,\uv} + k_i)^2
+ 2 \left(\frac{q_i}{2}- k_i - q_{12,\uv} \right) \cdot q_i  \right) \nn \\
&-& 4 \, (G_F(q_{12,\uv}))^2 \, \left( \left(q_{12,\uv} \cdot \left(q_i - k_i\right) \right)^2 + \frac{\mu_{\uv}^4}{d-2} \right)~.
\label{newselfUVscalartad}
\eea
The terms ${\bf \Delta}^{(1,\phi)}_{\Sigma, \uv}$ and ${\bf \Delta}^{(1,\phi)}_{T, \uv}$ integrate to zero independently in $d$ dimensions. In the latter, i.e. \Eq{newselfUVscalartad}, it was necessary to include a subleading contribution, proportional to $\mu^4_\uv$. We will not provide the integrated and dual expressions for ${\cal A}_{\Sigma,\uv}^{(2,f)} (q_i,q_{12,\uv})$ because they are quite heavy.

Finally, the counter-term ${\cal A}_{12,\uv}^{(2,f)}$ is simply obtained by considering the photon vertex and self-energy contributions together, namely
\bea
{\cal A}_{12,\uv}^{(2,t)}(q_i,q_{12,\uv})&=&{\cal A}_{\gamma,\uv}^{(2,t)} (q_i,q_{12,\uv})+{\cal A}_{\Sigma,\uv}^{(2,t)} (q_i,q_{12,\uv})~,\nn\\
{\cal A}_{12,\uv}^{(2,\phi)}(q_i,q_{12,\uv})&=&{\cal A}_{\gamma,\uv}^{(2,\phi)} (q_i,q_{12,\uv})+{\cal A}_{\gamma\gamma,\uv}^{(2,\phi)} (q_i,q_{12,\uv})+{\cal A}_{\Sigma,\uv}^{(2,\phi)} (q_i,q_{12,\uv})~.
\eea


\subsection{The double UV counter-term}\label{sec:doubleUV}
While it is entirely possible to compute ${\cal A}^{(2,f)}_{\uv^2}$ by directly taking the sum of all contributions, it is more interesting to consider well-chosen subsets of diagrams -- it also lightens intermediate expressions. For the top as the internal particle, it is logical to consider all the contributions to the Higgs boson vertex corrections, all the contributions to the photon vertex corrections and all the contributions to the self-energy corrections, as there is no ambiguity or cross-contributions for the $H\to\gamma\gamma$ process at two-loop. They will be written ${\cal A}^{(2,t)}_{H,\uv^2}$, ${\cal A}^{(2,t)}_{\gamma,\uv^2}$, ${\cal A}^{(2,t)}_{\Sigma,\uv^2}$, and account for 2, 4 and 6 diagrams, respectively. For the charged scalar as an internal particle, there is a subtlety. The three diagrams that contribute to the Higgs vertex correction also contribute to the $\gamma\gamma\phi\phi^\dagger$ vertex correction. For this reason, if we want to split ${\cal A}^{(2,\phi)}_{\uv^2}$, we have to consider both corrections together. For the photon and self-energy corrections, though, there is no ambiguity whatsoever. Thus, we define ${\cal A}^{(2,\phi)}_{H+\gamma\gamma,\uv^2}={\cal A}^{(2,\phi)}_{H,\uv^2}$, ${\cal A}^{(2,\phi)}_{\gamma,\uv^2}$, ${\cal A}^{(2,\phi)}_{\Sigma,\uv^2}$, that account for 9, 12 and 16 diagrams, respectively.\\

According to \Eq{AUV2}, the unintegrated double UV counter-terms have the form
\bea\label{DoubleUVForm}
{\cal A}^{(2,f)}_{\{H,\gamma,\Sigma\},\uv^2} &=& g_f \, s_{12} (e\, e_f)^2 \int_{\ell_1} \int_{\ell_{2}}\bigg[ (G_F(q_{1,\uv}))^{n_1} \, ( G_F(q_{2,\uv}))^{n_2} \, ( G_F(q_{12,\uv}) )^{n_{12}} \, \mathcal{N}_{\{H,\gamma,\Sigma\}}^{(f)}\nn\\
&-& 4\left(G_F(q_{1,\uv}) \right)^3\left(G_F(q_{12,\uv}) \right)^3d_{\{H,\gamma,\Sigma\},\uv^2}^{(f)} \, \mu_{\uv}^4 \bigg]~,
\eea
where
\beq
G_F(q_{2,\uv}) = \frac{1}{(q_{12,\uv}-q_{1,\uv})^2 - \mu_{\uv}^2 +\ii}~,
\eeq
and with $n_i$ being positive integers. Note that even though $\mathcal{N}_{\{H,\gamma,\Sigma\}}^{(f)}(q_{1,\uv},q_{12,\uv})$ should be expected to also depend on the external momenta $p_1$ and $p_2$, it is a remarkable feature that, thanks to welcome cancellations, it does not when considering the sum of all contributing diagrams. The expression in \Eq{DoubleUVForm} is therefore free of irreducible scalar products and can very easily be reduced through integrations by parts to the form
\beq
{\cal A}^{(2,f)}_{\{H,\gamma,\Sigma\},\uv^2}= g_f \, s_{12}(e\, e_f)^2 \left( c_{\{H,\gamma,\Sigma\},\ominus}^{(f)}\, I_\ominus + c_{\{H,\gamma,\Sigma\},\odot}^{(f)} \, I_\odot^2 \right),
\label{AHgammaSigmaUV2}
\eeq
where
\beq\label{MISunrise}
I_\ominus= \frac{1}{\mu_\uv^2} \, \int_{\ell_1}\,\int_{\ell_2}\, G_F(q_{1,\uv}, q_{12, \uv}, q_{2,\uv})=\left(\frac{\Se}{16\pi^2}\right)^2\left( \frac{\mu_\uv^2}{\mu^2}\right)^{-2\ep} \,
\left(-\frac{3}{2\epsilon^2}-\frac{9}{2\epsilon}+K_\ominus+\mathcal{O}(\epsilon)\right)~,
 \eeq
with
\beq
K_\ominus=-\frac{21}{2}+2\sqrt{3}\text{Cl}_2\left(\frac{\pi}{3}\right)~,
\eeq
where $\text{Cl}_2$ is the Clausen function of order 2, is the sunrise scalar integral with missing external momenta and all the internal masses equal, and	
\begin{equation}\label{MITadpole}
I_\odot = \frac{1}{\mu_\uv^2} \, \int_{\ell_i} G_F(q_{i,\uv}) = \frac{\Se}{16\pi^2}\left( \frac{\mu_\uv^2}{\mu^2}\right)^{-\ep} \frac{1}{\ep(1-\ep)}
\end{equation}
is the massive tadpole. Because of the presence of the double pole inside \Eq{MISunrise}, it is, in the general case, necessary to keep track of the different normalisations as choosing one over another could lead to a shift in the finite part. As we are working in the $\overline{\text{MS}}$ scheme, we should rather factorise the usual $S_\ep^{\overline{\text{MS}}}=(4\pi)^\epsilon e^{-\epsilon\,\gamma_E}$, but doing so, a global factor equal to $\Se/S_\ep^{\overline{\text{MS}}}=1+\delta S_\epsilon=1+\pi^2\,\epsilon^2/12+\mathcal{O}(\epsilon^3)$ would appear. We will see that in the end considering one normalisation over the other does not introduce any mismatch, so we chose to keep $\Se$ for simplicity.

These two master integrals (MI) are the only ones needed to evaluate ${\cal A}^{(2,f)}_{\uv^2}$ and fix the subleading terms $d_{\uv^2}^{(f)}$. Note that in \Eq{AHgammaSigmaUV2}, the dependence in $d_{\{H,\gamma,\Sigma\},\uv^2}^{(f)}$ is implicitly included in $c_{\{H,\gamma,\Sigma\},\odot}^{(f)}$~.\\

The unintegrated double UV counter-term for the diagrams with loop corrections in the Higgs boson vertex and
internal top quarks reads
\bea
{\cal A}_{H,\uv^2}^{(2,t)} &=& g_f \, s_{12} \,(e\, e_t)^2\, \int_{\ell_1} \int_{\ell_{2}} (G_F(q_{1,\uv}))^2  \, ( G_F(q_{12,\uv}) )^2 \bigg[-4\frac{d-4}{d-2}\left(c_{H,\uv}^{(t)}-G_F(q_{1,\uv})d_{H,\uv}^{(t)}\,\mu^2\right)\nn\\
&+&\frac{4G_F(q_{2,\uv})}{d-2}\left(d(d-4)q_{1,\uv}^2-2(d-2)^2q_{1,\uv}\cdot q_{12,\uv}+4\mu^2\right)\nn\\
&-&4G_F(q_{1,\uv})G_F(q_{12,\uv})d_{H,\uv^2}^{(t)}\,\mu^4\bigg]~,
\label{AHUV2}
\eea
and gives
\beq
c_{H,\ominus}^{(t)} = -\frac{2(d-3)((d-2)^2+4)}{3(d-2)}~, \qquad c_{H,\odot}^{(t)} = (d-2)\left(4+\frac{(d-4)^2}{4}\left(d_{H,\uv}^{(t)}-\frac{d-2}{4}d_{H,\uv^2}^{(t)}\right)\right)~.
\eeq
where the parameter $c_{H,\uv}^{(t)}$ has been replaced by its value, given in table~\ref{tab:renorm}. After integration, we have
\beq
\mathcal{A}^{(2,t)}_{H,\uv^2} = g_f\,s_{12}(e\,e_t)^2\,\left(\frac{\Se}{16\pi^2}\right)^2\left( \frac{\mu_\uv^2}{\mu^2}\right)^{-2\ep}\left(52+\frac{16K_\ominus}{3}+2d_{H,\uv}^{(t)}-d_{H,\uv^2}^{(t)}+\mathcal{O}(\epsilon)\right)~.
\eeq
The unintegrated expressions for the double UV counter-term for the photon and self-energy corrections are a bit heavy, so we will only provide the MI coefficients. For the photon vertex corrections, they read
\beq
c_{\gamma,\ominus}^{(t)} = \frac{8d(d-4)(d-2)}{3(d-2)}~, \qquad c_{\gamma,\odot}^{(t)} = -\frac{(d-4)(d-2)}{2}\left(2d-(d-4)d_{\gamma,\uv}^{(t)}+\frac{(d-4)(d-2)}{8}d_{\gamma,\uv^2}^{(t)}\right)~,
\eeq
while for the self-energy corrections, they read
\beq
c_{\Sigma,\ominus}^{(t)} = -\frac{4(d-4)^2(d-3)}{3(d-2)}~, \qquad c_{\Sigma,\odot}^{(t)} = \frac{(d-4)^2(d-2)}{4}\left(\frac{(d-4)(d-2)}{2}+d_{H,\uv}^{(t)}-4d_{\gamma,\uv}^{(t)}+4-\frac{d-2}{4}d_{\Sigma,\uv^2}^{(t)}\right)~,
\eeq
where once again we replaced $c_{H,\uv}^{(t)}$ and $c_{\gamma,\uv}^{(t)}$ by their value. Integrating the counter-terms gives
\beq
\mathcal{A}^{(2,t)}_{\gamma,\uv^2} = g_f\,s_{12}\,(e\, e_t)^2\left(\frac{\Se}{16\pi^2}\right)^2\left( \frac{\mu_\uv^2}{\mu^2}\right)^{-2\ep}\left(-16+4d_{\gamma,\uv}^{(t)}-d_{\gamma,\uv^2}^{(t)}+\mathcal{O}(\epsilon)\right)~,
\eeq
and
\beq
\mathcal{A}^{(2,t)}_{\Sigma,\uv^2} = g_f\,s_{12}\,(e\, e_t)^2\left(\frac{\Se}{16\pi^2}\right)^2\left( \frac{\mu_\uv^2}{\mu^2}\right)^{-2\ep}\left(4+2d_{H,\uv}^{(t)}-8d_{\gamma,\uv}^{(t)}-d_{\Sigma,\uv^2}^{(t)}+\mathcal{O}(\epsilon)\right)~,
\eeq
leading to
\beq
\mathcal{A}^{(2,t)}_{\uv^2} = g_f\,s_{12}\,(e\, e_t)^2\left(\frac{\Se}{16\pi^2}\right)^2\left( \frac{\mu_\uv^2}{\mu^2}\right)^{-2\ep}\left(40+\frac{16K_\ominus}{3}+4(d_{H,\uv}^{(t)}-d_{\gamma,\uv}^{(t)})-d_{\uv^2}^{(t)}+\mathcal{O}(\epsilon)\right)~,
\eeq
where we used $d_{H,\uv^2}^{(t)}+d_{\gamma,\uv^2}^{(t)}+d_{\Sigma,\uv^2}^{(t)}=d_{\uv^2}^{(t)}$.\\

For the scalar, the unintegrated double UV counter-term for the Higgs and $\gamma\gamma\phi\phi^\dagger$ corrections read
\bea
{\cal A}_{H+\gamma\gamma,\uv^2}^{(\phi)}&=&g_f\,s_{12}\,(e\, e_\phi)^2\int_{\ell_ 1}\,\int_{\ell_2}\,(G_F(q_{1,\uv}))^2  \, ( G_F(q_{12,\uv}) )^2 \bigg[2\frac{d-4}{d-2}c_{H,\uv}^{(\phi)}-\frac{c_{\gamma,\uv}^{(\phi)}}{d-2}\left(4-3d+4\mu^2G_F(q_{12,\uv})\right)\nn\\
&-&\frac{2G_F(q_{2,\uv})}{d-2}\left((d-4)q_{1,\uv}^2+(4-3d)q_{12,\uv}^2+2d\,q_{1,\uv}\cdot q_{12,\uv}+d\,\mu^2\right)\nn\\
&-&4G_F(q_{1,\uv})G_F(q_{12,\uv})d_{H+\gamma\gamma,\uv^2}^{(\phi)}\,\mu^4\bigg]~,
\eea
which after IBP reduction gives
\beq
c_{H+\gamma\gamma,\ominus}^{(\phi)}=0~, \qquad c_{\gamma,\odot}^{(\phi)}=-(d-2)\left(2+\frac{(d-4)^2(d-2)}{16}d_{H+\gamma\gamma,\uv^2}^{(\phi)}\right)~.
\eeq
We obtain
\beq
{\cal A}_{H+\gamma\gamma,\uv^2}^{(2,\phi)}=g_f\,s_{12}\,(e\, e_\phi)^2\left(\frac{\Se}{16\pi^2}\right)^2\left(-\frac{4}{\epsilon^2}-\frac{4}{\epsilon}-4-d_{H+\gamma\gamma,\uv^2}^{(\phi)}+\mathcal{O}(\epsilon)\right)~.
\eeq
As for the top, the unintegrated expressions for ${\cal A}^{(2,\phi)}_{\gamma,\uv^2}$ and ${\cal A}^{(2,\phi)}_{\Sigma,\uv^2}$ are a bit heavy. Their corresponding MI coefficients are
\beq
c_{\gamma,\ominus}^{(\phi)}=0~, \qquad c_{\gamma,\odot}^{(\phi)}=-(d-2)\left(4-\frac{(d-4)^2(d-2)}{16}d_{\gamma,\uv^2}^{(\phi)}\right)~.
\eeq
and
\beq
c_{\Sigma,\ominus}^{(\phi)}=-\frac{8(d-6)(d-3)}{3(d-2)}~, \qquad c_{\Sigma,\odot}^{(\phi)}=-\frac{d-2}{6}\left((d-6)\left((d-4)^2(d-2)-32\right)+(d-4)^2(d-2)d_{\Sigma,\uv^2}^{(\phi)}\right)~.
\eeq
After integration, we find
\beq
\mathcal{A}^{(2,\phi)}_{\gamma,\uv^2} = g_f\,s_{12}\,(e\, e_\phi)^2\left(\frac{\Se}{16\pi^2}\right)^2\left(\frac{8}{\epsilon^2}+\frac{8}{\epsilon}+8-d_{\gamma,\uv^2}^{(\phi)}+\mathcal{O}(\epsilon)\right)~,
\eeq
and
\beq
\mathcal{A}^{(2,\phi)}_{\Sigma,\uv^2} = g_f\,s_{12}\,(e\, e_\phi)^2\left(\frac{\Se}{16\pi^2}\right)^2\left(-\frac{4}{\epsilon^2}-\frac{4}{\epsilon}-22-\frac{8K_\ominus}{3}-d_{\Sigma,\uv^2}^{(\phi)}+\mathcal{O}(\epsilon)\right)~,
\eeq
leading to
\beq
\mathcal{A}^{(2,\phi)}_{\uv^2} = g_f\,s_{12}\,(e\, e_\phi)^2\left(\frac{\Se}{16\pi^2}\right)^2\left(-18-\frac{8K_\ominus}{3}-d_{\uv^2}^{(\phi)}+\mathcal{O}(\epsilon)\right)~,
\eeq
The coefficients $d_{\{H,\gamma,\Sigma\},\uv^2}^{(f)}$ can now be very easily be adjusted to obtain the desired $\mathcal{O}(\epsilon^0)$ part. Their value in the $\overline{\text{MS}}$ scheme are listed in table~\ref{tab:dUV2}.
\begin{table}[t]
\begin{center}
\begin{tabular}{|c|ccc|c|}
\hline
& $d_{H,\uv^2}^{(f)}$ & $d_{\gamma, \uv^2}^{(f)}$ & $d_{\Sigma, \uv^2}^{(f)}$ & $d_{\uv^2}^{(f)}$\\
\hline \hline
$t \bar{t}$ & $60+16K_\ominus/3$ & $-8~~~$ & $-4~~~$ & $48+16K_\ominus/3$ \\
$\phi\phi^\dagger$ & $-4~~~$ & $8$ & $-22-8K_\ominus/3$ & $-18-8K_\ominus/3~~~$ \\
\hline
\end{tabular}
\caption{Values of the scheme fixing paremeters in the $\overline{\text{MS}}$ for the double UV counter-terms.
\label{tab:dUV2}}
\end{center}
\end{table}

It is remarkable that for both particles, the full double UV counter-terms does not exhibit any $\epsilon$-poles, justifying in the mean time that using $\Se$ instead of $S_\epsilon^{\overline{\text{MS}}}$ does not introduce any discrepancy in the final result. While for the top quark each intermediate double UV counter-terms are all finite, for the charged scalar the cancellation of divergences only occurs when taking into account the sum of all 37 diagrams. This is due to the fact there are more subtle interplays between different contributing topologies, because of the more complex gauge structure. In both cases, the absence of divergences after integration means that in the traditional approach, these double UV counter-terms should not be needed. In our formalism however, they are essential in order to cancel local divergent behaviours appearing inside the amplitude and the single UV counter-terms. Note also that the unintegrated counter-terms exhibit terms proportional to $d-4$ that vanish when taking the four-dimensional limit at integrand level, even though they still lead to finite parts when keeping the $d$ dependence, because they multiply quantities that exhibit $\epsilon$-poles. This means that while the counter-terms $\mathcal{A}^{(2,f)}_{\uv^2}$ are finite, they will not lead to the same result if computed in $d$ or 4 dimensions. It is only when considering the renormalised amplitude $\mathcal{A}^{(2,f)}_{\r}$ that we have a strict and rigorous commutativity between integrating and taking the limit $\epsilon\to0$. The same happens at one-loop level \cite{Driencourt-Mangin:2017gop}, where the counter-term $A_{\uv}^{(1,f)}$ would not be needed (it integrates to 0 in the $\overline{\text{MS}}$) in $d$ dimensions. One can wonder if this property holds at three-loop order and beyond, if it is exclusive to the $H\to\gamma\gamma$ process and why, and if there is an underlying reason behind it.

\section{Numerical integration}
\label{sec:numint}

The renormalised unintegrated amplitude ${\cal A}^{(2,f)}_{\r}=\mathcal{A}^{(2,f)}-{\cal A}_{1,\uv}^{(2,f)}-{\cal A}_{12,\uv}^{(2,f)}-\mathcal{A}^{(2,f)}_{\uv^2} $ is completely free of local UV (and IR) singularities and can safely be evaluated in four dimensions. In the centre-of-mass frame of the decaying Higgs boson, we parametrise the three-momenta as
\begin{align}
\boldsymbol{\ell}_1&=\frac{\sqrt{s_{12}}}{2}\,\xi_1\,(\sin(\theta_1)\,\sin(\varphi_1),\sin(\theta_1)\,\cos(\varphi_1),\cos(\theta_1))~,\qquad&\mathbf{p}_1&=\frac{\sqrt{s_{12}}}{2}\,(0,0,1)\nn~,\\
\boldsymbol{\ell}_2&=\frac{\sqrt{s_{12}}}{2}\,\xi_2\,(\sin(\theta_2)\,\sin(\varphi_2),\sin(\theta_2)\,\cos(\varphi_2),\cos(\theta_2))~,\qquad&
\mathbf{p}_2&=\frac{\sqrt{s_{12}}}{2}\,(0,0,-1)~.
\end{align}
Note that we can assume $\boldsymbol{\ell}_1$, for instance, to belong to the $(x,z)$ plane, as there is a global rotational symmetry; this means we can trivially perform the integration over one azimuthal angle -- leading to a factor $2\pi$ -- and set $\varphi_1$ to 0. Furthermore, it is better to use $\boldsymbol{\ell}_{12}$ instead of $\boldsymbol{\ell}_2$ as integration variable, because of the way the UV counter-terms have been defined.
We therefore integrate over the two three-momenta
\begin{equation}
\boldsymbol{\ell}_1=\frac{\sqrt{s_{12}}}{2}\,\xi_1\,(\sin(\theta_1), 0 ,\cos(\theta_1))~,\quad\boldsymbol{\ell}_{12}=\frac{\sqrt{s_{12}}}{2}\,\xi_{12}\,(\sin(\theta_{12})\,\cos(\varphi_{12}),\sin(\theta_{12})\,\sin(\varphi_{12}),\cos(\theta_{12}))~,\\
\end{equation}
which leads to five integration variables, namely $\xi_1,\xi_{12}\in[0,\infty)$, $\theta_1,\theta_{12}\in[0,\pi]$ and $\varphi_{12}\in[0,2\pi]$. In addition, the usual compactification of the integration domain is performed, where the domains of $\xi_1$ and $\xi_{12}$ are mapped from $[0,\infty)$ onto to $[0,1]$, thanks to the change of variables
\begin{equation}\label{COVNum}
\xi_i\to\frac{x_i}{1-x_i}~,
\end{equation}
with $x_i\in[0,1]$, and $i\in\{1,12\}$. This increases the stability of the numerical integration for very high energies, because it restricts the local cancellation of UV singularities to a compact region. The integration measure, after applying this change of variables and in four dimensions, reads
\begin{equation}\label{IntegrationMeasure}
\frac{d\boldsymbol{\ell}_1\,d\boldsymbol{\ell}_2}{(2\pi)^8}=\frac{s_{12}^3}{64}\frac{x_1^2\,x_{12}^2\,\sin(\theta_1)\sin(\theta_{12})}{(1-x_1)^4(1-x_{12})^4}\frac{dx_1\,dx_{12}\,d\theta_1\,d\theta_{12}\,d\varphi_{12}}{(2\pi)^7}~.
\end{equation}

Directly writing all the dual cuts explicitly in terms of the integration variables would not lead to a reasonable computational time, as the integration would be too heavy numerically speaking. Moreover, it would be far from being optimal, since many identical terms would have to be evaluated more than once. Instead, we take advantage of the fact it was possible to write all dual cuts in terms of the reduced denominators $D_i$, as explained in section \ref{sec:reduction} and shown in appendix \ref{ap:unrenormalised}. The first step, which only needs to be done once, is to express all the reduced denominators in terms of the 5 integration variables, and this for each dual cut\footnote{Recall that the expression of a given $D_i$ differs from dual cut to dual cut.}. We can then compute their numerical values for a given point in the integration domain, which allows us to quickly evaluate the integrand at this very point by appropriately replacing each $D_i$. Thus, regardless of the complexity of a given integrand, only a limited amount of objects have to be numerically evaluated. Although quite simple to implement, this strategy helped decreasing the integration time by more than one order of magnitude. To perform the actual integration, we used the in-built \mathematica~function \verb"NIntegrate". Our results are shown in figure \ref{fig:plots}, where they are compared with the analytical results given in appendix~\ref{TwoLoopAnalytical}. The integration time is of a few minutes for each point. The biggest source of numerical error comes from the cancellation, between the amplitude and the double UV counter-term, of the non-decoupling term going as $\mathcal{O}(M_f^2/s_{12})$. For $M_t^2=2s_{12}$ for example, the part being removed from the amplitude by the counter-term is two orders of magnitude higher than the actual result, effectively multiplying the relative numerical error by a factor 100, roughly. Note that this error is twice as big for the top quark as for the charged scalar, because of the relative factor -2 between their respective $\mathcal{O}(M_f^2/s_{12})$ terms. Nevertheless, the agreement with the analytical result is excellent for all values of the internal and renormalisation masses considered. Numerical instabilities may however appear when considering $r$ very close to 4 (or equivalently $M_f$ very close to $\sqrt{s_{12}}/2$), as we approach the mass threshold. Beyond this limit ($r>4$), contour deformation would be needed. Within the LTD formalism, numerical implementations of contour deformation have already been successfully implemented in \cite{Buchta:2015wna}.
\begin{figure}[t]
\begin{center}
\includegraphics[width=0.45\textwidth]{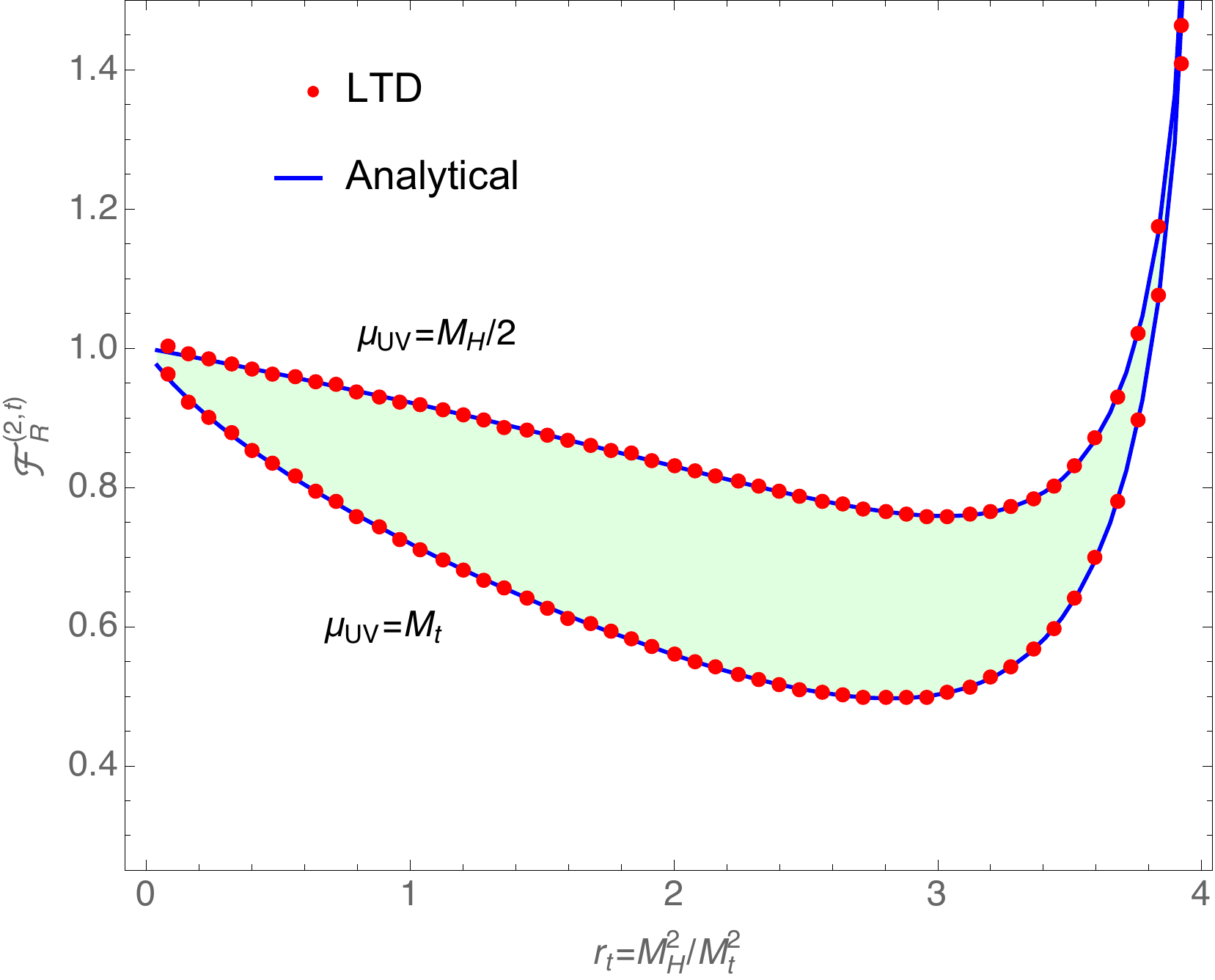} 
$\qquad$
\includegraphics[width=0.462\textwidth]{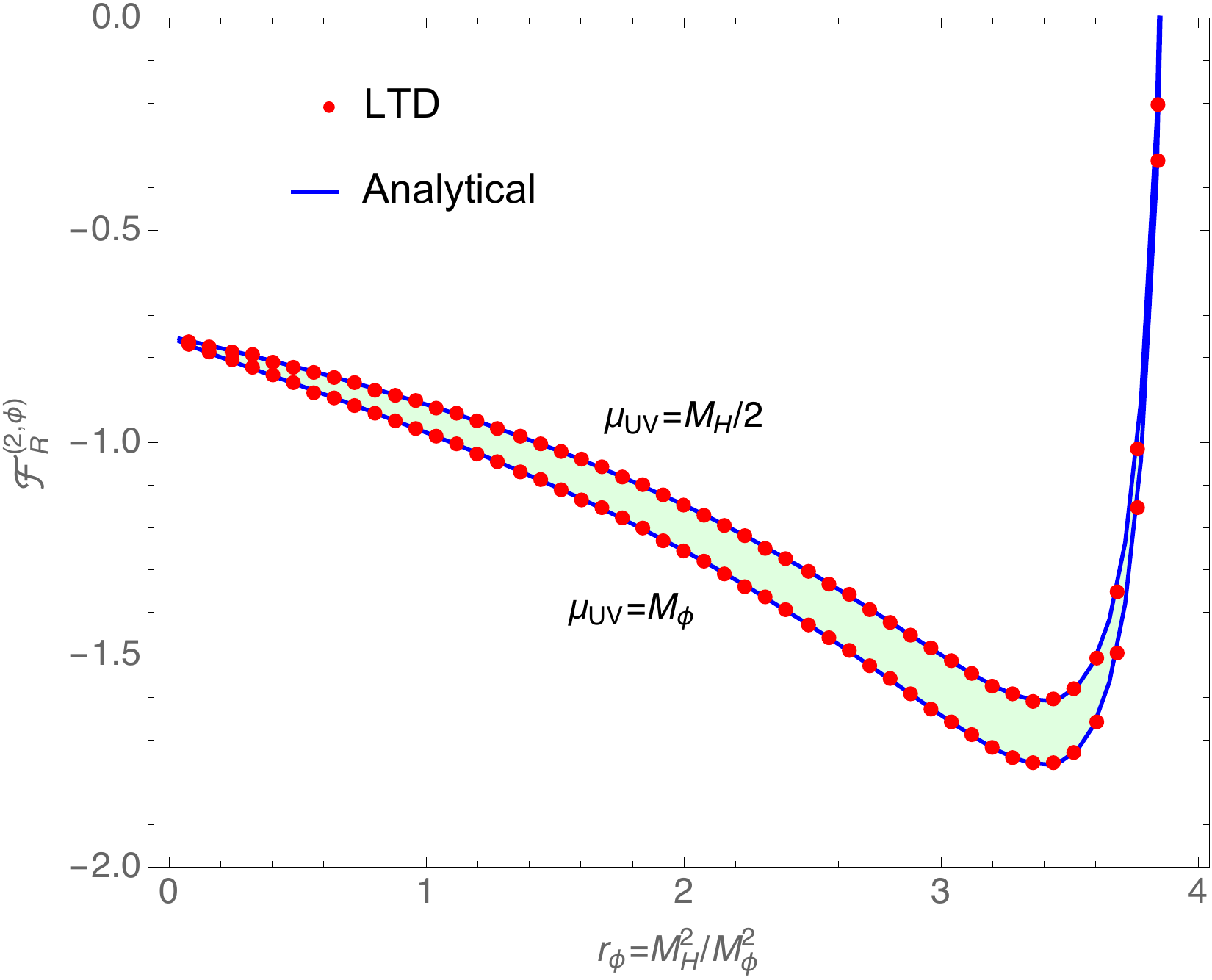}
\caption{Integrated renormalised amplitude of the two-loop corrections to the $H\to\gamma\gamma$ process, as a function of the inverse mass square of the particle running inside the loop. On the left we show the top quark contribution and on the right, the charged scalar contribution. The solid blue lines represent the analytical result using DREG, while the red dots have been obtained numerically with the LTD formalism. For each particle, two values of the renormalisation scale have been considered, namely $\mu_\uv=M_H/2=\sqrt{s_{12}}/2$, and $\mu_\uv=M_f$. We have rescaled our results as we used different normalisations. Explicitly, $\mathcal{F}_\r^{(2,f)}=64\pi^6\,{\cal A}^{(2,f)}_{\r}$.\label{fig:plots}}
\end{center}
\end{figure}
\section{Conclusion}
\label{sec:conclusions}
In this paper, we have studied a purely four-dimensional representation of the renormalised Higgs boson decay amplitudes at two-loop order, so that it can directly be evaluated numerically. For this purpose, we have applied the Loop-Tree Duality (LTD) theorem to the $H \to \gamma \gamma$ process, where internal particles are charged scalars and fermions. Working at the two-loop level within this formalism involves performing double-cuts and several algebraic manipulations of the expressions. This has been done through a fully automatised \mathematica~code which can be adapted to deal with any two-loop scattering amplitude.

Once the full sets of double-cuts were generated, we have taken advantage of the previously known one-loop results \cite{Driencourt-Mangin:2017gop}, in order to infer the universal integrand-level structure of the two-loop expressions. Surprisingly, we have found many similarities between both cases, which has allowed us to write the full two-loop amplitude at integrand level using the same functional forms independently of the nature of the particles circulating inside the loop. As in the one-loop case, the explicit process dependence is codified inside specific scalar coefficients. 

We have kept the $d$ dependence in all intermediate steps, although a noticeable simplification of the universal coefficients takes place in the limit $d=4$. We have studied the singular structure of the two-loop amplitudes to understand how to achieve a purely four-dimensional representation of the finite parts. In the first place, we have studied the cancellation of spurious threshold singularities that appear as a consequence of splitting the different cuts. When considering $H \to \gamma \gamma$ below the physical threshold, the amplitude is guaranteed to be infrared safe as well as free of any physical threshold singularity. In fact, we have managed to prove that all the spurious singularities vanish in this configuration after putting together all possible double cuts.

Then, we have developed a fully local framework to remove ultraviolet singularities. We have implemented an algorithmic approach to renormalise, at integrand level, the two-loop amplitudes. It is based on the refinement of the expansion around the UV propagator strategy \cite{Driencourt-Mangin:2017gop,Sborlini:2016gbr,Sborlini:2016hat}, including an iteration to remove all the possible UV divergences. A careful study of the UV structure of vertices and self-energies has been performed, which allowed to impose constraints on the finite remainder containing the specific renormalisation scheme dependence. Furthermore, we have also shown that the UV scale we used in the derivation of the local counter-terms corresponds to the renormalisation scale used in the traditional approach. These last two points allowed to build the required counter-terms at integrand level to reproduce the $\overline{\text{MS}}$ results.

Finally, we have proceeded to combine the universal dual representations of the $H \to \gamma \gamma$ amplitudes together with the local UV counter-terms, achieving a fully renormalised integrand in four space-time dimensions. This has allowed a purely numerical implementation which fully agrees with the available results in the literature. On top of that, intermediate checks with the scalar sunrise diagram were implemented, in order to test the reliability of the codes.

The results presented in this paper constitutes a major advance into the extension of the Four-Dimensional Unsubtraction (FDU) formalism \cite{Sborlini:2016gbr,Sborlini:2016hat} to NNLO accuracy, paving the way for a more efficient and purely numerical implementation of currently relevant physical processes.

\section*{Acknowledgements}
This work is supported by the Spanish Government (Agencia Estatal de Investigaci\'on) and ERDF funds from European Commission (Grants No. FPA2017-84445-P and SEV-2014-0398), by Generalitat Valenciana (Grant No. PROMETEO/2017/053) and by Consejo Superior de Investigaciones Cient\'{\i}ficas (Grant No. PIE-201750E021). 
FDM acknowledges support from Generalitat Valenciana (GRISOLIA/2015/035) and GFRS from Fondazione Cariplo under the Grant No. 2015-0761. WJT acknowledges support from the Spanish government (FJCI-2017-32128). Also, FDM and GFRS were supported by a STSM Grant from the COST Action CA16201 PARTICLEFACE.

\appendix
\section{Unrenormalised two-loop dual amplitudes for $H\to \gamma\gamma$}
\label{ap:unrenormalised}

In this appendix, we collect the explicit expressions for the unrenormalised two-loop dual amplitudes for $H\to \gamma\gamma$.
The subindices take the values $i\in\{1,2\}$ and $j,k\in\{3,12\}$, with $j\ne k$. The functions $G$ and $F$ have been defined in \Eq{eq:FG}. 
Here, we introduce the auxiliary function 
\beq
H(X,\kappa) = - \frac{\kappa}{X} \frac{\partial X}{\partial \kappa}~.
\eeq
For example
\beq
H(D_3 \, D_{12}, \kappa_i) = \kappa_i\, b_{1,0} \left( \frac{1}{D_3}-\frac{1}{D_{12}} \right)~,
\eeq

\beq
H(D_1 \, D_k, \kappa_j) = \kappa_j\, b_{1,0} \left( \frac{1}{D_1}+\frac{2}{D_k} \right)~,
\eeq

\beq
H(D_k^2, \kappa_j) = -\kappa_j\, b_{1,0} \, \frac{4}{D_k}~,
\eeq
with $b_{1,0} = 2\, p_{1,0}/M_f$ (recall that $p_{1,0}=p_{2,0}$). It is very important to note that the permutation inside the following expressions have to be applied on the parameters of $H$, instead of the expression of $H$ itself after derivation, as the symmetry is not any more explicit.
\subsection{Double cuts from $G_D(\alpha_1) G_D(\alpha_2) G_F(\alpha_3)$}

This is the only set with direct snail contributions for scalars, the only terms that do not 
depend on $\ell_2$ are those proportional to $c_{4,u}$ and $c_{16}$. This subset generates 4 different double cuts that are obtained from the following expressions:
\bea
{\cal A}^{(2,f)}(q_i,q_4)&=&g_f^{(2)}\,\int_{\ell_1}\int_{\ell_2} \, \tilde{\delta}(q_i,q_4)\,\Bigg\{ - \frac{r_f\,c_1^{(f)}}{D_3\,D_{12}}\Bigg(G(D_{\overline{i}},\kappa_i,c_{4,u}^{(f)})\big(1+H(D_3\,D_{12},\kappa_i)\big)+F(D_{\overline{i}},\kappa_4/\kappa_i)\Bigg) \nn \\
&+& \Bigg( c_7^{(f)}\left(\frac{1}{D_{\overline{i}}} - \frac{1}{D_{\overline{3}}}\left(1-\frac{D_3}{D_{12}}\left(1-\frac{D_{\overline{12}}}{D_{\overline{i}}}\right)\right)\right)
\nn \\ &+& \frac{1}{D_3} \left( c_8^{(f)} \left(\frac{1}{D_{\overline{3}}}-\frac{1}{D_{\overline{i}}}\right) 
-\frac{1}{D_{\overline{12}}}\left(c_9^{(f)}-c_{10}^{(f)}\frac{D_{\overline{3}}}{D_{\overline{i}}}\right) \right)\nn \\
&+& 2\, r_f \, \Bigg[\frac{1}{D_3\,D_{12}}\left(c_1^{(f)}\left(\frac{1}{D_3\,D_{\overline{3}}}
+\frac{1}{D_{\overline{i}}}\left(\frac{1}{D_{\overline{3}}}-\frac{1}{D_3} \right)\right)
+\frac{c_{14}^{(f)}}{D_{\overline{3}}} + \frac{c_{20}^{(f)}}{D_{\overline{i}}}-c_{16}^{(f)} \right. \nn \\
&+& \left. c_{17}^{(f)}\left(\frac{D_{\overline{i}}-D_{\overline{12}}}{D_{\overline{3}}}+\frac{D_{\overline{3}}}{D_{\overline{i}}}\right)\right)-\frac{1}{D_{\overline{i}}\,D_{\overline{3}}}\left(\frac{c_7^{(f)}}{D_{12}}+c_{18}^{(f)} \right)\Bigg] +\{3\leftrightarrow12\}\Bigg)\Bigg\}~,	
\eea
and
\bea
\mathcal{A}^{(2,f)}(q_j,q_4)&=&g_f^{(2)}\,\int_{\ell_1}\int_{\ell_2}\tilde{\delta}(q_j,q_4)\Bigg\{-\frac{r_f}{D_k}\Bigg[\frac{c_1^{(f)}}{D_1}\Bigg(G(D_{\overline{j}},\kappa_j,c_{4,u}^{(f)})\big(1+H(D_1\,D_k,\kappa_j)\big)+F(D_{\overline{j}},\kappa_4/\kappa_j)\Bigg)\nonumber\\
&-&\frac{c_{23}^{(f)}}{2}\Bigg(G(D_{\overline{j}},\kappa_j,c_{4,u}^{(f)})\big(1+\frac{1}{2}H(D_k^2,\kappa_j)\big)+F(D_{\overline{j}},\kappa_4/\kappa_j)\Bigg)\Bigg]\nn\\
&+& c_7^{(f)}\left(\frac{1}{D_{\overline{k}}}\left(\frac{D_{\overline{j}}}{D_{\overline{1}}}+\frac{D_k}{D_1}\left(1-\frac{D_{\overline{j}}}{D_{\overline{1}}}\right)\right)-\frac{1}{D_{\overline{1}}}\right)+
\frac{1}{D_1} \left( c_8^{(f)}\left(\frac{1}{D_{\overline{j}}}-\frac{1}{D_{\overline{1}}}\right)-\frac{c_9^{(f)}}{D_{\overline{k}}} \right) \nn\\
&+& \frac{1}{D_{\overline{1}}} \left(
c_{10}^{(f)}\left(\frac{D_{\overline{j}}}{D_{\overline{k}}}\left(\frac{1}{D_1}-\frac{1}{D_k}\right)-\frac{D_{\overline{k}}}{D_k\, D_{\overline{j}}}\right)+\frac{2c_{11}^{(f)}}{D_k} 
+\frac{c_{13}^{(f)}}{D_{\overline{k}}}
\right)+\frac{c_{12}^{(f)}}{D_k}\left(\frac{1}{D_{\overline{j}}}+\frac{1}{D_{\overline{k}}}\right) \nn\\
&+&2\, r_f\Bigg[\frac{1}{D_k}\Bigg(c_1^{(f)}\left(\frac{1}{D_1}\left(\frac{1}{D_{\overline{j}}}\left(\frac{1}{D_{\overline{1}}}-\frac{1}{D_k}-\frac{1}{D_1}-\frac{1}{2}\right)+\frac{1}{D_{\overline{1}}\,D_{\overline{k}}}\left(1+\frac{D_{\overline{k}}}{D_1}+\frac{D_{\overline{1}}}{D_k}\right)\right) \right.\nn\\
&+&\left. \frac{2-r_f}{2\, D_{\overline{1}}\,D_{\overline{j}}\,D_{\overline{k}}}\right)
- \frac{2c_{16}^{(f)}}{D_1}+c_{17}^{(f)}\left(\frac{1}{D_1}\left(\frac{D_{\overline{j}}+D_{\overline{k}}}{D_{\overline{1}}}-\frac{D_{\overline{k}}}{D_{\overline{j}}}-\frac{D_{\overline{j}}}{D_{\overline{k}}}\right) \right.\nn \\
&+&\left. \left(\frac{D_1}{D_{\overline{1}}}+\frac{D_{\overline{1}}}{D_1}\right)\left(\frac{1}{D_{\overline{j}}}+\frac{1}{D_{\overline{k}}}\right)+\frac{c_7^{(f)}\, r_f}{D_{\overline{j}}\,D_{\overline{k}}}\left(1-\frac{D_1}{D_{\overline{1}}}\right)\right)\nonumber\\
&+&\left(\frac{c_{14}^{(f)}}{D_1}+\frac{c_{19}^{(f)}}{D_{\overline{1}}}\right)\left(\frac{1}{D_{\overline{j}}}+\frac{1}{D_{\overline{k}}}\right)\Bigg)+\frac{1}{D_{\overline{1}}}\left(\frac{c_{18}^{(f)}}{D_{\overline{j}}}-\frac{2c_{15}^{(f)}}{D_1\,D_k}-\frac{1}{D_{\overline{k}}}\left(\frac{c_7^{(f)}}{D_1}-\frac{c_{20}^{(f)}}{D_{\overline{j}}}\right)\right)\nn\\
&-&\frac{c_{23}^{(f)}}{8}\left(\frac{1}{D_{\overline{j}}\,D_{\overline{k}}}+\frac{4}{D_k}\left(\frac{1}{D_k\,D_{\overline{k}}}-\frac{1}{D_{\overline{j}}}\left(\frac{1}{D_k}+\frac{1}{2}-\frac{2-r_f}{2\, D_{\overline{k}}}\right)\right)\right)\Bigg]\Bigg\}+\{1\leftrightarrow2\}~.
\eea

\subsection{Double cuts from $G_G(\alpha_1) G_D(-\alpha_2) G_D(\alpha_3)$}

There are also 4 double cuts in this subset that are obtained from the expressions:
\bea
{\cal A}^{(2,f)}(q_{\overline{4}},q_{\overline{i}})&=& g_f^{(2)}\,\int_{\ell_1}\int_{\ell_2} \, \td{q_{\overline{4}},q_{\overline{i}}}
\Bigg\{ - c_7^{(f)}\left(\frac{1}{D_3}\left(1-\frac{D_{\overline{3}}}{D_{\overline{12}}}\left(1-\frac{D_{12}}{D_i}\right)\right) - \frac{1}{D_i}\right) \nn \\ 
&+&\frac{1}{D_3}\left(-\frac{c_8^{(f)}}{D_i}+c_{10}^{(f)}\frac{D_{\overline{3}}}{D_{\overline{12}}}\left(\frac{1}{D_i}-\frac{1}{D_{12}}\right)
+\frac{c_{11}^{(f)}}{D_{12}}+\frac{c_{13}^{(f)}}{D_{\overline{12}}}\right)\nn \\ 
&+&2\, r_f \Bigg[\frac{1}{D_3\,D_{12}}\left(c_1^{(f)}\left(\frac{1}{D_i}\left(\frac{1}{2D_i}+\frac{1}{D_{\overline{3}}}\right)+\frac{2-r_f}{4 D_{\overline{3}}\,D_{\overline{12}}}\right)-\frac{c_{15}^{(f)}}{D_i} \right. \nn \\ 
&+& \left. c_{17}^{(f)} \left(\frac{D_i}{D_{\overline{12}}}+\frac{D_{\overline{12}}}{D_i}
- c_7^{(f)} \frac{r_f\,  D_i}{2 D_{\overline{3}} D_{\overline{12}}} 
\right) +\frac{c_{19}^{(f)}}{D_{\overline{12}}} 
\right) \nn \\
&+&\frac{1}{D_{\overline{12}}}\left(\frac{1}{D_3}\left(-\frac{c_7^{(f)}}{D_i}+\frac{c_{20}^{(f)}}{D_{\overline{3}}}\right)+c_{18}^{(f)}\left(\frac{1}{D_{12}}-\frac{1}{D_i}\right)\right)\Bigg]\Bigg\} + \{3\leftrightarrow12\}~,
\eea
and
\bea
{\cal A}^{(2,f)}(q_{\overline{4}},q_{\overline{j}})&=& g_f^{(2)}\,\int_{\ell_1}\int_{\ell_2} \, \td{q_{\overline{4}},q_{\overline{j}}} \Bigg\{
c_7^{(f)}\left(\frac{1}{D_k}\left(\frac{D_j}{D_1}+\frac{D_{\overline{k}}}{D_{\overline{1}}}\left(1-\frac{D_j}{D_1}\right)\right)-\frac{1}{D_1}\right) \nn \\
&+& \frac{c_8^{(f)}}{D_1 D_j} 
+\frac{1}{D_k}\left(-  \frac{c_9^{(f)}}{D_1} + 
c_{10}^{(f)}\, \frac{D_{\overline{k}}}{D_{\overline{1}}} \left(\frac{1}{D_1}-\frac{1}{D_j}\right)+\frac{c_{12}^{(f)}}{D_j}+\frac{c_{13}^{(f)}}{D_{\overline{1}}}\right) \nn \\ 
&+&2\, r_f\Bigg[\frac{1}{D_j\,D_k}\left(c_1^{(f)}\left(\frac{1}{D_1}\left(\frac{1}{D_j}+\frac{1}{D_{\overline{1}}}\right)+\frac{2-r_f}{2 D_{\overline{1}}\,D_{\overline{k}}}\right) + \frac{c_{14}^{(f)}}{D_1} \right. \nn \\ 
&+& \left. c_{17}^{(f)} \left( \frac{D_1}{D_{\overline{1}}}+\frac{D_{\overline{1}}-D_{\overline{k}}}{D_1}
+ c_{7}^{(f)} \, \frac{r_f}{D_{\overline{k}}} \left(  1-\frac{D_1}{D_{\overline{1}}} \right) 
\right) + \frac{c_{19}^{(f)}}{D_{\overline{1}}} \right) \nn\\
&+&  \frac{1}{D_{\overline{1}}}\left(-\frac{c_7^{(f)}}{D_1\,D_k}+\frac{c_{20}^{(f)}}{D_{\overline{k}}}\left(\frac{1}{D_j}+\frac{1}{D_k}\right)+c_{18}^{(f)}\left(\frac{1}{D_j}-\frac{1}{D_1} \right) \right) \nn\\
&-& \frac{c_{23}^{(f)}}{8} \left(\frac{1}{D_{\overline{k}}}\left(\frac{1}{D_j}+\frac{1}{D_k} \right) + \frac{4}{D_j\,D_k} \left(
\frac{1}{D_j} + \frac{2-r_f}{2\, D_{\overline{k}}} \right) \right)  
\Bigg]\Bigg\} + \{1\leftrightarrow2\}~.
\eea

\subsection{Double cuts from $G_D(\alpha_1) G_F(\alpha_2) G_D(\alpha_3)$}

In this subset there are 14 double cuts. The terms that do not depend on $D_4$ or $D_{\overline{k}}$ integrate to a massive snail. The generating expressions are:
\bea
{\cal A}^{(2,f)}(q_i,q_{\overline{i}})&=& g_f^{(2)}\,\int_{\ell_1}\int_{\ell_2} \, \td{q_i,q_{\overline{i}}} \nn \\ &\times&\Bigg\{
- \frac{r_f \,c_1^{(f)}}{D_3\,D_{12}}\Bigg(G(D_4,\kappa_i,-c_{4,nu}^{(f)})\big(1+H(D_3\,D_{12},\kappa_i)\big)+F(D_4,-\kappa_{\overline{i}}/\kappa_i)\Bigg) \nn \\
&+& \Bigg( c_7^{(f)}\left(\frac{1}{D_4}\left(1-\frac{D_{12}\,D_{\overline{3}}}{D_3\,D_{\overline{12}}}\right)-\frac{4\, c_{4,nu}^{(f)}}{D_3}\left(1-\frac{D_{\overline{3}}}{D_{\overline{12}}}\right)\right)-\frac{1}{D_3\,D_4}\left(c_8^{(f)}-c_{10}^{(f)}\frac{D_{\overline{3}}}{D_{\overline{12}}}\right) \nn\\
&+&2\, r_f \Bigg[\frac{1}{D_3\,D_{12}}\Bigg(c_1^{(f)}\left(\frac{1}{D_4}\left(\frac{1}{D_{\overline{3}}}-\frac{1}{D_3}\right)+\frac{c_{4,nu}^{(f)}}{D_3}\right)+\frac{1}{D_4}\left(c_{20}^{(f)}+c_{17}^{(f)}\,D_{\overline{3}}\right)+\frac{c_{21}^{(f)}}{D_{\overline{3}}}\Bigg) \nn\\
&+&\frac{c_7^{(f)}}{D_{12}}\left(\frac{1}{D_{\overline{3}}}\left(\frac{1}{2}-\frac{1}{D_4}\right)+\frac{2\, c_{4,nu}^{(f)}}{D_3}\left(1-\frac{D_4}{D_{\overline{3}}}\right)\right)-\frac{c_{18}^{(f)}}{D_4\,D_{\overline{3}}}\Bigg]+\{3\leftrightarrow12\} \Bigg)\Bigg\}~,
\eea
\bea
{\cal A}^{(2,t)}(q_j,q_{\overline{j}})&=&g_f^{(2)}\,\int_{\ell_1}\int_{\ell_2}\tilde{\delta}(q_j,q_{\overline{j}}) \nn \\
&\times& \Bigg\{-\frac{r_f}{D_k}\Bigg[\frac{c_1^{(f)}}{D_1}\Bigg(G(D_4,\kappa_j,-c_{4,nu}^{(f)})\big(1+H(D_1\,D_k,\kappa_j)\big)+F(D_4,-\kappa_{\overline{j}}/\kappa_j)\Bigg) \nn\\
&-&\frac{c_{23}^{(f)}}{2}\Bigg(G(D_{\overline{j}},\kappa_j,-c_{4,nu}^{(f)})\big(1+\frac{1}{2}H(D_k^2,\kappa_j)\big)+F(D_4,-\kappa_{\overline{j}}/\kappa_j)\Bigg)\Bigg] \nn \\
&+& 4\, c_7^{(f)}\,c_{4,nu}^{(f)}\left(\frac{1}{D_1}-\frac{D_{\overline{k}}}{D_{\overline{1}}\, D_k}\right)+\frac{1}{D_4}\left(\frac{c_8^{(f)}}{D_1}-\frac{1}{D_k}\left(c_{10}^{(f)}\frac{D_{\overline{k}}}{D_{\overline{1}}}-c_{12}^{(f)}\right)\right)\nonumber\\
&+&2\, r_f \Bigg[\frac{1}{D_k}\left(c_1^{(f)}\left(\frac{1}{D_4}\left(\frac{1}{D_1}\left(\frac{1}{D_{\overline{1}}}-\frac{1}{D_1}-\frac{1}{D_k}-\frac{1}{2}\right)+\frac{2-r_f}{2D_{\overline{1}}\,D_{\overline{k}}}\right)+\frac{c_{4,nu}^{(f)}}{D_1}\left(\frac{1}{D_1}+\frac{1}{D_k}\right)\right)\right.\nonumber\\
&+&\frac{1}{D_4} \left( \frac{c_{14}^{(f)}}{D_1} + c_{17}^{(f)} \left(\frac{D_1}{D_{\overline{1}}}+\frac{D_{\overline{1}}-D_{{\overline{k}}}}{D_1}
+\frac{c_7^{(f)}\, r_f}{D_{\overline{k}}} \left(1-\frac{D_{\overline{1}}}{D_1} \right)\right)\right) \nn \\ 
&+&2\, c_7^{(f)}\,c_{4,nu}^{(f)}\left(\frac{1}{D_1}+\frac{1}{D_{\overline{1}}}\left(1-\frac{D_4}{D_1}\right)\right)
+ \left.\frac{1}{D_{\overline{1}}}\left(\frac{c_{19}^{(f)}}{D_4}+\frac{c_{21}^{(f)}}{D_1}+\frac{c_{22}^{(f)}}{D_{\overline{k}}}\right)\right)+\frac{1}{D_4\,D_{\overline{1}}}\left(c_{18}^{(f)}+\frac{c_{20}^{(f)}}{D_{\overline{k}}}\right)\nn\\
&-&\frac{c_{23}^{(f)}}{8}\left( \frac{1}{D_4}\left( \frac{1}{D_{\overline{k}}}
- \frac{4}{D_k}\left(\frac{1}{D_k}+\frac{1}{2}-\frac{2-r_f}{2D_{\overline{k}}}\right)\right)
+ \frac{4\, c_{4,nu}^{(f)}}{D_k}\left(\frac{1}{D_k}-\frac{1}{D_{\overline{k}}}\right)\right) \Bigg]\Bigg) \Bigg\}+\{1\leftrightarrow2\}~, \nn \\
\eea
\bea
{\cal A}^{(2,f)}(q_i,q_{\overline{j}})&=& g_f^{(2)}\,\int_{\ell_1}\int_{\ell_2} \, \td{q_i,q_{\overline{j}}} \Bigg\{
-c_7^{(f)} \left( \frac{1}{D_4} - \frac{4c_{4,nu}^{(f)}}{D_j} \right)\left(1 - \frac{D_j}{D_k}\left(1-\frac{D_{\overline{k}}}{D_{\overline{i}}}\right)\right) \nn \\
&+& \frac{1}{D_4}\left(\frac{c_8^{(f)}}{D_j}-\frac{c_9^{(f)}}{D_k}+c_{10}^{(f)}\frac{D_{\overline{k}}}{D_k\,D_{\overline{i}}}\right) \nn \\
&+& 2\, r_f \Bigg[\frac{1}{D_j\,D_k}\left(c_1^{(f)}\left(\frac{1}{D_4}\left(\frac{1}{D_j}+\frac{1}{D_{\overline{i}}}\right)-\frac{c_{4,nu}^{(f)}}{D_j}\right)
+\frac{1}{D_4}\left(c_{14}^{(f)}+c_{17}^{(f)}(D_{\overline{i}}-D_{\overline{k}})\right)+\frac{c_{21}^{(f)}}{D_{\overline{i}}}\right)\nn \\ 
&+&\frac{c_7^{(f)}}{D_k}\left(\frac{1}{D_{\overline{i}}}\left(\frac{1}{2}-\frac{1}{D_4} \right)+\frac{2c_{4,nu}}{D_j}\left(1-\frac{D_4}{D_{\overline{i}}}\right)\right)-\frac{c_{18}^{(f)}}{D_4\,D_{\overline{i}}}\Bigg]\Bigg\}~,
\eea
\bea
{\cal A}^{(2,f)}(q_j,q_{\overline{i}})&=& g_f^{(2)}\,\int_{\ell_1}\int_{\ell_2} \, \td{q_j,q_{\overline{i}}}\Bigg\{-c_7^{(f)}\left(
\left( \frac{1}{D_4} - \frac{4c_{4,nu}^{(f)}}{D_k}\right)
\left(1-\frac{D_{\overline{j}}}{D_{\overline{k}}}\left(1-\frac{D_k}{D_i}\right)\right)\right. \nn \\
&+&\left.4c_{4,nu}^{(f)}\left( \frac{1}{D_i} - \frac{1}{D_k} \left( 1- \frac{D_{\overline{k}}}{D_{\overline{j}}}\right) \right) \right) \nn\\
&+&\frac{1}{D_4}\left(-\frac{c_8^{(f)}}{D_i}+c_{10}^{(f)}\left(\frac{D_{\overline{j}}}{D_{\overline{k}}}\left(\frac{1}{D_i}-\frac{1}{D_k}\right)-\frac{D_{\overline{k}}}{D_k\,D_{\overline{j}}}\right)+\frac{2c_{11}^{(f)}}{D_k}+\frac{c_{13}^{(f)}}{D_{\overline{k}}}\right) \nn\\
&+&2\, r_f\Bigg[\frac{1}{D_k} \left(c_1^{(f)} \left( \frac{1}{D_4} \left(
\frac{1}{D_i}\left(\frac{1}{D_i}+\frac{1}{D_{\overline{j}}}+\frac{1}{D_{\overline{k}}}\right)+\frac{2-r_f}{2 D_{\overline{j}}\,D_{\overline{k}}}\right)
-\frac{c_{4,nu}^{(f)}}{D_i^2} \right)
+\frac{c_{21}^{(f)}}{D_i}\left(\frac{1}{D_{\overline{j}}}+\frac{1}{D_{\overline{k}}}\right)\right. \nn\\
&+&\left.\frac{c_{22}^{(f)}}{D_{\overline{j}}\,D_{\overline{k}}}\right)+ c_7^{(f)} \left(\frac{1}{D_i\,D_{\overline{k}}}\left(\frac{1}{2}-\frac{1}{D_4}\right)+
\frac{2c_{4,nu}^{(f)}}{D_k}\left( \frac{2}{D_i}+ \left(1-\frac{D_4}{D_i}\right)\left(\frac{1}{D_{\overline{j}}}+\frac{1}{D_{\overline{k}}}\right)\right)\right) \nn\\
&+&\frac{1}{D_4}\left(\frac{1}{D_{\overline{j}}}\left(c_{18}^{(f)}+\frac{c_{20}^{(f)}}{D_{\overline{k}}}\right)+\frac{1}{D_k}\left(-\frac{2c_{15}^{(f)}}{D_i}+c_{17}^{(f)}\left(D_i\left(\frac{1}{D_{\overline{j}}}+\frac{1}{D_{\overline{k}}}  -c_7^{(f)}\,\frac{r_f}{D_{\overline{j}}\,D_{\overline{k}}}\right) \right. \right. \right.\nn \\
&+& \left.\left. \left. \frac{D_{\overline{j}}+D_{\overline{k}}}{D_i}\right)
+c_{19}^{(f)}\left(\frac{1}{D_{\overline{j}}}+\frac{1}{D_{\overline{k}}}\right)\right)\right)\Bigg]\Bigg\}~,
\eea
and
\bea
A^{(2,t)}(q_j,q_{\overline{k}})&=&g_f^{(2)}\,\int_{\ell_1}\int_{\ell_2} \,\Bigg\{c_7^{(f)}
\left( \frac{1}{D_4} - \frac{4 c_{4,nu}^{(f)}}{D_k}\right) 
\left(\frac{D_{\overline{j}}}{D_{\overline{1}}}+\frac{D_k}{D_1}\left(1-\frac{D_{\overline{j}}}{D_{\overline{1}}}\right)\right)\nonumber\\
&+&\frac{1}{D_4}\left(-\frac{c_9^{(f)}}{D_1}+c_{10}^{(f)}\frac{D_{\overline{j}}}{D_{\overline{1}}}\left(\frac{1}{D_1}-\frac{1}{D_k}\right)+\frac{c_{12}^{(f)}}{D_{12}}+\frac{c_{13}^{(f)}}{D_{\overline{1}}}\right)\nn\\
&+&2\, r_f\Bigg[\frac{1}{D_k}\left(c_1^{(f)}\left(\frac{1}{D_4}\left(\frac{1}{D_1}\left(\frac{1}{D_k}+\frac{1}{D_{\overline{1}}}\right)+\frac{2-r_f}{2D_{\overline{1}}\,D_{\overline{j}}}\right)-\frac{c_{4,nu}^{(f)}}{D_1\,D_k}\right)\right.\nonumber\\
&+&\left.\frac{1}{D_4}\left(\frac{c_{14}^{(f)}}{D_1}+c_{17}^{(f)}\left(\frac{D_1}{D_{\overline{1}}}+\frac{D_{\overline{1}}-D_{\overline{j}}}{D_1}
+c_7^{(f)}\, \frac{r_f}{D_{\overline{j}}}\left(1-\frac{D_1}{D_{\overline{1}}}\right)\right)+\frac{c_{19}^{(f)}}{D_{\overline{1}}}\right)\right)\nn\\
&+&c_7^{(f)}\left(\frac{1}{D_1\,D_{\overline{1}}}\left(\frac{1}{2}-\frac{1}{D_4}\right)+\frac{2 c_{4,nu}^{(f)}}{D_k}\left(\frac{1}{D_1}+\frac{1}{D_{\overline{1}}}\left(1-\frac{D_4}{D_1}\right)\right)\right) \nn \\
&+&\frac{1}{D_{\overline{1}}}\left(\frac{c_{20}^{(f)}}{D_{\overline{j}}\,D_4}+\frac{c_{21}^{(f)}}{D_1\,D_k}+\frac{c_{22}^{(f)}}{D_k\,D_{\overline{j}}}\right)\nonumber\\
&-&\frac{c_{23}^{(f)}}{8}\left( \frac{1}{D_4}\left(\frac{1}{D_{\overline{j}}}+
\frac{4}{D_k}\left(\frac{1}{D_k}+\frac{2-r_f}{2 D_{\overline{j}}}\right)\right)-\frac{4c_{4,nu}^{(f)}}{D_k}\left(\frac{1}{D_k}+\frac{1}{D_{\overline{j}}}\right)\right)\Bigg]\Bigg\}\nn\\
&+&\{1\leftrightarrow2\}~.
\eea

\section{Known analytic results for $H\to \gamma\gamma$ at two loops}
\label{TwoLoopAnalytical}
In this appendix, we write the analytical results obtained from ref.~\cite{Aglietti:2006tp}, for the top quark and the charged scalar in the $\overline{\text{MS}}$ scheme. We first define
\begin{equation}
x_f=\frac{\sqrt{1-4/r_f}-1}{\sqrt{1-4/r_f}+1}
\end{equation}
where we recall $r_f = s_{12}/M_f^2$. We also write the auxiliary function
\bea
\mathcal{H}_1(x)&=&\frac{9}{10}\zeta_2^2+2\zeta_3\,H(0,x)+\zeta_2\,H(0,0,x)+\frac{1}{4}H(0,0,0,0,x)+\frac{7}{2}H(0,1,0,0,x)\nn\\
&-&2H(0,-1,0,0,x)+4H(0,0,-1,0,x)-H(0,0,1,0,x)~,
\eea
where the standard harmonic polylogarithm notations~\cite{Remiddi:1999ew} have been used. For the top quark,
\bea
\mathcal{F}_\r^{(2,t)}(x)&=&\frac{36x}{(x-1)^2}-\frac{4x(1-14x+x^2)}{(x-1)^4}\zeta_3-\frac{4x(1+x)}{(x-1)^3}H(0,x)-\frac{8x(1+9x+x^2)}{(x-1)^4}H(0,0,x)\nn\\
&+&\frac{2x(3+25x-7x^2+3x^3)}{(x-1)^5}H(0,0,0,x)\nn\\
&+&\frac{4x(1+2x+x^2)}{(x-1^4)}\big(\zeta_2\,H(0,x)+4H(0,-1,0,x)-H(0,1,0,x)\big)\nn\\
&+&\frac{4x(5-6x+5x^2)}{(x-1)^4}H(1,0,0,x)-\frac{8x(1+x+x^2+x^3)}{(x-1)^5}\mathcal{H}_1(x)\nn\\
&-&\left(\frac{12x}{(x-1)^2}-\frac{6x(1+x)}{(x-1)^3}H(0,x)+\frac{6x(1+6x+x^2)}{(x-1)^4}H(0,0,x)\right)\log\left(\frac{M_t^2}{\mu_\uv^2}\right)~,
\eea
whereas for the charged scalar,
\bea
\mathcal{F}_\r^{(2,\phi)}(x)&=&-\frac{14x}{(x-1)^2}-\frac{24x^2}{(x-1)^4}\zeta_3+\frac{x(3-8x+3x^2)}{(x-1)^3(x+1)}H(0,x)+\frac{34x^2}{(x-1)^4}H(0,0,x)\nn\\
&-&\frac{8x^2}{(x-1)^4}\big(\zeta_2\,H(0,x)+4H(0,-1,0,x)-H(0,1,0,x)+H(1,0,0,x)\big)\nn\\
&-&\frac{2x^2(5-11x)}{(x-1)^5}H(0,0,0,x)+\frac{16x^2(1+x^2)}{(x-1)^5(x+1)}\mathcal{H}_1(x)\nn\\
&+&\left(\frac{6x^2}{(x-1)^3(x+1)}H(0,x)-\frac{6x^2}{(x-1)^4}H(0,0,x)-\frac{3}{4}\mathcal{F}_\r^{(1,\phi)}(x)\right)\log\left(\frac{M_\phi^2}{\mu_\uv^2}\right)~.
\eea
with
\beq
\mathcal{F}_\r^{(1,\phi)}(x)=\frac{4}{r_\phi}\left(1+\frac{2}{r_\phi}H(0,0,x)\right)~.
\eeq


\bibliographystyle{JHEP}
\bibliography{refs}

\end{document}